%% file: MAXIJ1820.tex
\documentclass[fleqn,usenatbib]{mnras}

\DeclareRobustCommand{\VAN}[3]{#2}
\let\VANthebibliography\thebibliography
\def\thebibliography{\DeclareRobustCommand{\VAN}[3]{##3}\VANthebibliography}
\DeclareRobustCommand{\DE}[3]{#2}
\let\DEthebibliography\thebibliography
\def\thebibliography{\DeclareRobustCommand{\DE}[3]{##3}\DEthebibliography}

\usepackage[T1]{fontenc}
\usepackage{graphicx}
\usepackage{amsmath}
\usepackage{amssymb}
\usepackage{placeins}
\usepackage{lineno}
\usepackage{soul}
\usepackage{float}
\usepackage{newtxtext,newtxmath}

\title[Gamma-ray observations of MAXI~J1820+070]{Gamma-ray observations of MAXI~J1820+070 during the 2018 outburst}

\input{author_list}

\date{Accepted XXX; Received YYY; in original form ZZZ}
\pubyear{2022}

\begin{document}
\label{firstpage}
\pagerange{\pageref{firstpage}--\pageref{lastpage}}
\maketitle

\pagebreak

\begin{abstract}

MAXI~J1820+070 is a low-mass X-ray binary with a black hole as a compact object. This binary underwent an exceptionally bright X-ray outburst from March to October 2018, showing evidence of a non-thermal particle population through its radio emission during this whole period. The combined results of 59.5~hours of observations of the MAXI~J1820+070 outburst with the H.E.S.S., MAGIC and VERITAS experiments at energies above 200~GeV are presented, together with \textit{Fermi}-LAT data between 0.1 and 500~GeV, and multiwavelength observations from radio to X-rays. Gamma-ray emission is not detected from MAXI~J1820+070, but the obtained upper limits and the multiwavelength data allow us to put meaningful constraints on the source properties under reasonable assumptions regarding the non-thermal particle population and the jet synchrotron spectrum. In particular, it is possible to show that, if a high-energy gamma-ray emitting region is present during the hard state of the source, its predicted flux should be at most a factor of 20 below the obtained \textit{Fermi}-LAT upper limits, and closer to them for magnetic fields significantly below equipartition. During the state transitions, under the plausible assumption that electrons are accelerated up to $\sim 500$~GeV, the multiwavelength data and the gamma-ray upper limits lead consistently to the conclusion that a potential high-energy and very-high-energy gamma-ray emitting region should be located at a distance from the black hole ranging between $10^{11}$ and $10^{13}$~cm. Similar outbursts from low-mass X-ray binaries might be detectable in the near future with upcoming instruments such as CTA.

\end{abstract}

\begin{keywords}
stars: individual: MAXI~J1820+070 -- gamma rays: general -- stars: black holes -- X-rays: binaries
\vspace{-12mm}
\end{keywords}


\section{Introduction}\label{sec:introduction}

X-ray binaries are systems in which a compact object -- either a black hole (BH) or a neutron star -- accretes matter from a companion star. In low-mass X-ray binaries, the companion mass is below $\sim 1~M_\odot$, and accretion on to the compact object normally takes place through an accretion disk generated by the Roche lobe overflow mechanism \citep[e.g.][]{remillard06}. Typically, low-mass X-ray binaries with a BH (BH-LMXBs) also feature transient jets launched from the BH, which are powered by the accretion process, the magnetic field, the BH rotation, or a combination of them \citep[see][and references therein]{romero17}. These jets can efficiently accelerate charged particles, potentially up to GeV or TeV energies, and emit non-thermal radiation from radio to gamma rays as a result of the radiative cooling of the accelerated particles \citep[see e.g.,][for a review on jets in X-ray binaries]{mirabel99,fender16}.

Most of the time, BH-LMXBs are in a quiescent state until they undergo periodic outbursts likely triggered by variations in the properties of the accretion disk that result in a change of the mass accretion rate on to the BH \citep[e.g.][]{fender12}. During one of these outbursts, that may last for several months, the luminosity of a BH-LMXB increases by several orders of magnitude. A BH-LMXB can be detected in a soft state (SS) or a hard state (HS) based on the hardness of its X-ray spectrum during one of these outbursts. At the beginning of the outburst, a BH-LMXB is typically in the HS, in which the X-rays exhibit a hard-spectrum component. This emission likely originates in a hot corona around the BH, where inverse Compton (IC) scattering of low-energy photons coming from the accretion disk takes place. The HS also features jet synchrotron emission, which is mostly seen at radio and infrared wavelengths, although it may also be responsible for a significant contribution to the X-ray output of the system \citep[e.g.][]{fender16}. As the outburst continues, the source will transition to the SS. In this state, most of the X-rays are of thermal origin, emitted by the hot inner regions of the accretion disk. Also, radio emission fades away, indicating a lack of jet activity (although weak jets may still be present and remain undetected). In a typical outburst, a BH-LMXB normally completes the HS--SS--HS cycle, going through short-lived intermediate states during the HS--SS and SS--HS transitions. As happened with the triggering of the outburst, the changes in the spectral states of BH-LMXBs are probably produced by variations in the accretion disk properties. During the state transitions, especially the HS--SS one, discrete blobs of plasma moving away from the BH can sometimes be resolved in radio, rather than the continuous jets typical of the HS \citep[see][and references therein for a more detailed description of the states of BH-LMXBs]{fender12}.

With one possible exception, no high-energy (HE, above 100~MeV) or very-high-energy (VHE, above 100~GeV) gamma-ray emission is detected from BH-LMXBs \citep{magic17_v404,hess18_mq}. The possible exception to this is the $\sim 4\sigma$ excess at HE of V404 Cygni during an outburst in 2015 (\citealt{loh16,piano17}; although we note the lack of a significant excess reported by a recent reanalysis of the \textit{Fermi}-LAT data; \citealt{harvey21}). A firm detection of BH-LMXBs at HE or VHE would enable a better physical characterisation of these systems in terms of their magnetic field, particle acceleration mechanisms and maximum particle energy, or gamma-ray absorption processes, among others. We note that LMXBs hosting a neutron star have been detected at HE \citep[see, e.g.,][]{harvey22}. In these systems, the gamma-ray emission likely originates in processes involving the neutron star, which are therefore not applicable in a BH scenario \citep[e.g.,][and references therein]{strader16}.

For high-mass X-ray binaries, there is already evidence for gamma-ray emission. HE gamma rays are detected from systems like Cygnus~X-1 and Cygnus~X-3, likely originating from the jets in both cases \citep[see][]{zanin16,fermi09,agile09,zdziarski18}. HE emission is also detected from regions of SS433 far from the central binary, where the jets terminate interacting with the supernova remnant around the source \citep{fang20,li20}. On the other hand, the VHE detection of high-mass X-ray binaries is still elusive \citep{magic10_cygx3,magic15_mwc656,magic17_cygx1,veritas_cygx3,veritas_v404_4U0115}, with the exception of SS433 (and excluding gamma-ray binaries from this source class). For this source, the HAWC Collaboration detected photons with energies of $\sim 20$~TeV originating in regions very far from the binary system, although not spatially coincident with the HE-emitting sites. The post-trial detection significance ranged from $4.0\sigma$ to $4.6\sigma$ depending on the analysed region, and reached $5.4\sigma$ for a joint fit of the interaction regions \citep{hawc18_ss433}.

MAXI~J1820+070 (RA = $18^{\rm h} 20^{\rm m} 21\fs9$, Dec = $+07\degr 11\arcmin 07\arcsec$; Galactic coordinates $l$ = $35.8536\degr$, $b$ = $+10.1592\degr$) is a BH-LMXB discovered in the optical band on 2018 March 6 (MJD~58184.1) by the All-Sky Automated Survey for Supernovae (ASAS-SN; \citealt{tucker18}), and on March 11 (MJD~58188.5) was also detected in X-rays by the Monitor of All-sky X-ray Image (MAXI; \citealt{kawamuro18}). Soon after its discovery, MAXI~J1820+070 showed an exceptionally high X-ray flux peaking at $\sim 4$ times that of the Crab Nebula \citep[e.g.,][]{delsanto18,shidatsu19}. A distance to the source of $d = 2.96 \pm 0.33$~kpc was determined from radio parallax \citep{atri20}, which is consistent with the distance of $3.28^{+0.60}_{-0.52}$~kpc obtained from \textit{Gaia} DR3 data \citep{bailer-jones21}. Jet activity was detected from MAXI~J1820+070 in the form of radio and infrared emission, which classifies the source as a microquasar \citep[e.g.,][]{bright20,rodi21}. The Lorentz factor of the jet during the HS was estimated to be $\Gamma = 1.7 - 4.1$ from radio-to-optical data, the upper and lower limits of this range being determined from constraints on the jet power and the pair production rate, respectively \citep{zdziarski22_jet}. For discrete ejections taking place in the HS--SS transition, a Lorentz factor of $\Gamma = 2.2^{+2.8}_{-0.5}$ was obtained \citep[using a distance to the source of 2.96~kpc;][]{atri20}. The jet inclination was measured to be $\theta = 64^\circ \pm 5^\circ$ from radio observations \citep{wood21}, and its half-opening angle in the HS was found to be $1.3^\circ \pm 0.7^\circ$ \citep{zdziarski22_jet}. Using optical polarisation observations, the jet misalignment with respect to the perpendicular to the orbital plane was measured to be at least $40^\circ$ \citep[with a $68\%$ confidence level;][]{poutanen22}. An orbital period of $16.4518 \pm 0.0002$~h was determined from optical spectroscopic observations \citep{torres19}. The BH and stellar masses were constrained through further spectroscopy measurements to a 95\% confidence interval of $5.7 - 8.3~M_\odot$ and $0.28 - 0.77~M_\odot$, respectively, for orbital inclinations between $66^\circ$ and $81^\circ$ \citep{torres20}. The parameters above yield an orbital semi-major axis of $\sim 4.5\times10^{11}$~cm. An estimate of the donor star parameters is discussed in \cite{2022ApJ...930....9M}.

\begin{figure*}
	\centering
    \includegraphics[width=\textwidth]{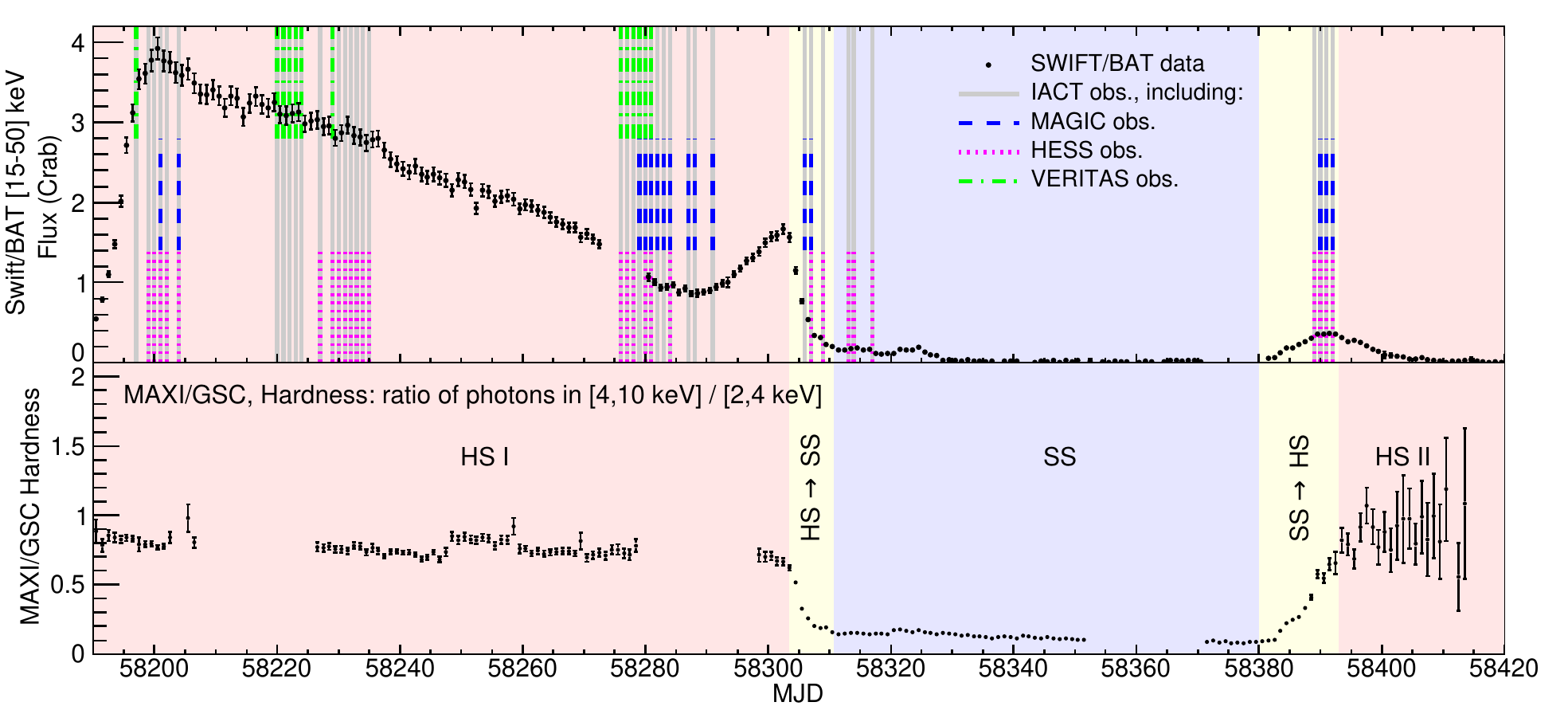}
    \caption{\textit{Top panel}: X-ray flux in the 15--50 keV band as seen by \textit{Swift}/BAT, with the VHE observation dates superimposed as vertical lines with different styles for each collaboration. Only the days with data after quality cuts are shown. \textit{Bottom panel}: Evolution of the MAXI~J1820+070 hardness ratio of the 4--10 to 2--4~keV fluxes as seen by MAXI/GSC. The source states are superimposed as light red (HS), light blue (SS) and light yellow (HS--SS / SS--HS) background colours.}
    \label{fig:HR}
\end{figure*}

MAXI~J1820+070 remained in the HS from the beginning of the outburst in March until early July (2018), when it began its transition to the SS. This source state lasted until late September, when MAXI~J1820+070 started transitioning back to the HS shortly before becoming quiescent and putting an end to the outburst, which lasted a total of $\sim 7$ months. During its outburst, MAXI~J1820+070 was observed with a wide variety of instruments at radio \citep[e.g.,][]{atri20,bright20}, near infrared \citep[e.g.][]{sanchezsierras20}, optical \citep[e.g.,][]{veledina19,torres19,shidatsu19}, and X-ray \citep[e.g.,][]{roques19,shidatsu19,buisson19,fabian20,chakraborty20,zdziarski21_acc} frequencies. We make use of the results of \cite{shidatsu19} to define the exact dates of the beginning and end of each source state, based on the MAXI Gas Slit Camera (MAXI/GSC) hardness ratio (i.e., the flux ratio of high-energy to low-energy X-rays) between the $6-20$~keV and $2-6$~keV photon fluxes. These dates are shown in Table~\ref{tab:states}. The evolution of the X-ray state of MAXI~J1820+070 can be seen in the bottom panel of Fig.~\ref{fig:HR}, which shows its hardness ratio from MAXI/GSC data.

\begin{table}
    \begin{center}
    \caption{Starting and ending times used for each X-ray state of MAXI~J1820+070, based on the results of \protect\cite{shidatsu19}. Hard State I and II refer, respectively, to the initial and final states of the source, as depicted in Fig.~\ref{fig:HR}.}
    \begin{tabular}{l c c c c}
    \hline \hline
    Source state            & Start     & End       & Start         & End           \\
                            & [MJD]     & [MJD]     & [Gregorian]   & [Gregorian]   \\
    \hline
    Hard State I            & 58189.0   & 58303.5   & 12 Mar. 2018  & 4 Jul. 2018   \\
    HS $\rightarrow$ SS     & 58303.5   & 58310.7   & 4 Jul. 2018   & 11 Jul. 2018  \\
    Soft State              & 58310.7   & 58380.0   & 11 Jul. 2018  & 19 Sep. 2018  \\
    SS $\rightarrow$ HS     & 58380.0   & 58393.0   & 19 Sep. 2018  & 2 Oct. 2018   \\
    Hard State II           & 58393.0   & 58420.0   & 2 Oct. 2018   & 29 Oct. 2018  \\
    \hline
    \end{tabular}
    \label{tab:states}
    \end{center}
\end{table}

In this work, we present the results of combined observations, by the H.E.S.S., MAGIC and VERITAS collaborations, of VHE gamma rays from MAXI~J1820+070, the brightest BH-LMXB in X-rays ever observed. In order to give a more complete picture of the source, \textit{Fermi}-LAT data in HE gamma rays are also included, as well as multiwavelength observations from radio to X-rays. This work is structured as follows: Section~\ref{sec:observations} describes the HE and VHE observations and data analysis for each telescope. Section~\ref{sec:results} presents the results of this work, for which a discussion is given in Section~\ref{sec:discussion}. Finally, we conclude with a summary in Section~\ref{sec:summary}.


\section{Observations and data analysis}\label{sec:observations}

\subsection{H.E.S.S., MAGIC and VERITAS data}\label{sec:VHE}

MAXI~J1820+070 was observed during its 2018 outburst with the H.E.S.S., MAGIC and VERITAS Imaging Atmospheric Cherenkov Telescope (IACT) arrays.
H.E.S.S. is an array of five IACTs located in the Khomas Highland, Namibia ($23^\circ$S, $16^\circ$E, 1800~m above sea level). It comprises four telescopes with a 12-m diameter dish and a Field of View (FoV) of $5^\circ$ \citep[for a description, see][]{2006A&A...457..899A}, and one telescope with a 28-m diameter dish and a $3.2^\circ$ FoV \citep{Bolmont_2014}. H.E.S.S. investigates gamma rays in the energy range from $\sim 20$~GeV \citep{2018A&A...620A..66H} to $\sim 100$~TeV \citep{2021A&A...653A.152A}.
MAGIC \citep{magic16a} is a stereoscopic system of two IACTs located at the Roque de los Muchachos Observatory in La Palma, Spain ($29^\circ$N, $18^\circ$W, 2200~m above sea level). The telescopes have a $3.5^\circ$ FoV, and are equipped with a primary dish with a diameter of 17~m. MAGIC can detect gamma rays from $\sim 15$~GeV \citep{magic20_geminga} to $\sim 100$~TeV \citep{magic20_vlza}.
VERITAS is an array of four IACTs located at the Fred Lawrence Whipple Observatory in southern Arizona, USA \citep[$32^\circ$N, $111^\circ$W, 1270~m above sea level;][]{weekes2002}. Each telescope covers a FoV of $3.5^\circ$, collecting light from a 12-m diameter reflector. VERITAS is sensitive to gamma-ray photons ranging from $\sim 85$~GeV to $\gtrsim 30$~TeV. The performance of VERITAS is described in \cite{park2015}.

The MAGIC and H.E.S.S. observations were performed from March to October 2018, covering the initial HS of the source, the beginning of the SS, and the state transitions. The VERITAS data were collected from March to June, when the source was in the HS. After data quality cuts, 26.3~h, 22.5~h and 10.7~h of effective observation time (defined as the exposure time corrected for dead-time losses) remains for H.E.S.S., MAGIC and VERITAS, respectively, for a combined total of 59.5~h. The data sample was divided according to the X-ray state (or transition) of the source as defined in Sect.~\ref{sec:introduction}. A summary of the observations, including their zenith angle, is shown in Table~\ref{tab:observations}. The observation dates of each telescope are shown in Table~\ref{tab:dates}, and they are also represented in Fig.~\ref{fig:HR} superimposed on the hard X-ray light curve (LC) of the source.

\begin{table}
    \begin{center}
    \caption{Summary of the observations of MAXI~J1820+070 by the H.E.S.S., MAGIC and VERITAS collaborations, after data quality cuts. The effective observation time, the zenith angle range and its median are shown for each source state and experiment.}
    \begin{tabular}{c c c c}
    \hline \hline
    Source state	        & Experiment    & Time [h]	& Zenith angle (median) [deg]   \\
    \hline
    Hard State              & H.E.S.S.      & 17.9		& 30 -- 61 (33)                 \\
                            & MAGIC         & 14.2		& 21 -- 58 (34)                 \\
                            & VERITAS       & 10.7		& 20 -- 39 (28)                 \\
    \hline
    HS $\rightarrow$ SS     & H.E.S.S.      & 4.0		& 30 -- 38 (32)                 \\
                            & MAGIC         & 4.9		& 21 -- 48 (27)                 \\
    \hline
    Soft State              & H.E.S.S.      & 2.6		& 30 -- 34 (31)                 \\
                            
    \hline
    SS $\rightarrow$ HS     & H.E.S.S.      & 1.8		& 37 -- 53 (43)                 \\
                            & MAGIC         & 3.4		& 28 -- 56 (41)                 \\
                            
    \hline
    TOTAL                   & H.E.S.S.      & 26.3		& 30 -- 61 (33)                 \\
                            & MAGIC         & 22.5		& 21 -- 58 (32)                 \\
                            & VERITAS       & 10.7		& 20 -- 39 (28)                 \\
    \hline
    \end{tabular}
    \label{tab:observations}
    \end{center}
\end{table}

The low-level data analyses of H.E.S.S., MAGIC and VERITAS were performed using standard collaboration procedures, each of them including an independent cross-check (i.e., an independent analysis, performed with a different software pipeline, that yielded compatible results with the main analysis). These low-level analyses comprise, among others, calibration and image cleaning procedures, methods to separate atmospheric showers triggered by gamma rays from those triggered by hadrons, and gamma-ray energy and direction reconstruction (see \citealt{de_naurois_2009,holler2015photon} for the H.E.S.S. main analysis; \citealt{2014APh....56...26P} for the H.E.S.S. cross-check; \citealt{magic16b} for MAGIC; \citealt{daniel2007} for the VERITAS main analysis; and \citealt{maier2017} for the VERITAS cross-check). The VHE emission was assumed point-like (since it is expected to come from regions close to the binary system, with angular sizes smaller than the instruments' resolutions), and the signal region was defined by a radius of $0.12^\circ$, $0.14^\circ$ or $0.10^\circ$ around the source position for H.E.S.S., MAGIC and VERITAS, respectively. In order to maximise the source effective observation time, and thus the probability of detection, a joint analysis of the data from the three experiments was also done (see Appendix~\ref{sec:fluxComputation}). No significant signal was detected from the individual or combined data sets, regardless of the energy range considered. The gamma-ray upper limits (ULs) in different energy bins were computed following a maximum-likelihood ratio test as described in Appendix~\ref{sec:fluxComputation}, both for the individual and combined data sets. We also refer the reader to \cite{magichess18_ss433} for a similar method. A confidence level (C.L.) of 0.95 was used, and a global flux systematic uncertainty of 30\% was taken for each experiment, which accounts for the systematic error in both the flux normalisation and the energy scale \citep[see e.g.][]{2012A&A...539L...2A}. The choice of a common value of the systematic uncertainty for the three experiments is motivated by the similar values of the systematic errors among them \citep[see][for the estimation of systematic errors in H.E.S.S., MAGIC and VERITAS, respectively]{2006A&A...457..899A,magic16b,2022A&A...658A..83A}. The VHE gamma-ray spectrum was assumed to follow a power law with spectral index $\alpha = 2.5$, i.e. ${\rm d}N/{\rm d}\varepsilon \propto \varepsilon^{-\alpha}$, where $N$ is the number of gamma-ray photons and $\varepsilon$ is their energy. This spectral shape is chosen as it resembles what has been observed for other binary systems detected at VHE \citep[e.g.][]{hess06_ls5039,magic06_lsi,vmh21_J0632}, since similar particle acceleration mechanisms and non-thermal emission processes of VHE gamma rays are expected to occur in MAXI~J1820+070.

\subsection{\textit{Fermi}-LAT data}\label{sec:Fermi}

The Large Area Telescope (LAT) \citep{atwood09} is a pair-conversion detector on the \textit{Fermi Gamma-Ray Space Telescope}. It consists of a tracker and a calorimeter, each of them made of a $4 \times 4$ array of modules, an anticoincidence detector that covers the tracker array, and a data acquisition system with a programmable trigger. The \textit{Fermi}-LAT is located at a low-Earth orbit with 90~min period and normally operates in survey mode, with a $2.4$~sr FoV. Such an observational strategy allows the instrument to cover the whole sky in approximately 3~h. The data selected for the analysis presented in this paper cover the period MJD 58189 -- 58420. The $0.1 - 500$~GeV data were analysed with the latest available \texttt{fermitools} v.~2.0.8 with P8R3\_V3 response functions (\texttt{SOURCE} photon class; maximum zenith angle of $90^\circ$).

A standard binned likelihood analysis \citep{mattox96} of the data taken from a $14^\circ$-radius region of interest (ROI) around the MAXI~J1820+070 position was performed\footnote{See e.g. \url{https://fermi.gsfc.nasa.gov/ssc/data/analysis/scitools/binnededisp_tutorial.html}}. The analysis is based on the fitting of a spatial and spectral model of the sky region around the source of interest to the data. The model of the region included all sources from the 4FGL DR3 catalogue~\citep{4fgl_catalogue} as well as components for isotropic and galactic diffuse emissions given by the standard spatial and spectral templates \texttt{iso\_P8R3\_SOURCE\_V3\_v1.txt} and \texttt{gll\_iem\_v07.fits}. 

The spectral template for each 4FGL source in the region was selected according to the catalogue model. The normalisations of the spectra of these sources, as well as the normalisations of the Galactic diffuse and isotropic backgrounds, were assumed to be free parameters during the fit. MAXI~J1820+070 was modelled as a point-like source with a power-law spectrum. Following the recommendation of the \textit{Fermi}-LAT collaboration, our analysis is performed with the energy dispersion handling enabled. To minimise the potential effects from the sources present beyond the considered ROI, we additionally included into the model all the 4FGL sources up to $10^\circ$ beyond the ROI, with all the spectral parameters fixed to the catalogue values. The parameters used for the \textit{Fermi}-LAT analysis are summarised in Table~\ref{tab:fermi_analysis_details}.

\begin{figure*}
	\centering
    \includegraphics[width=0.32\textwidth]{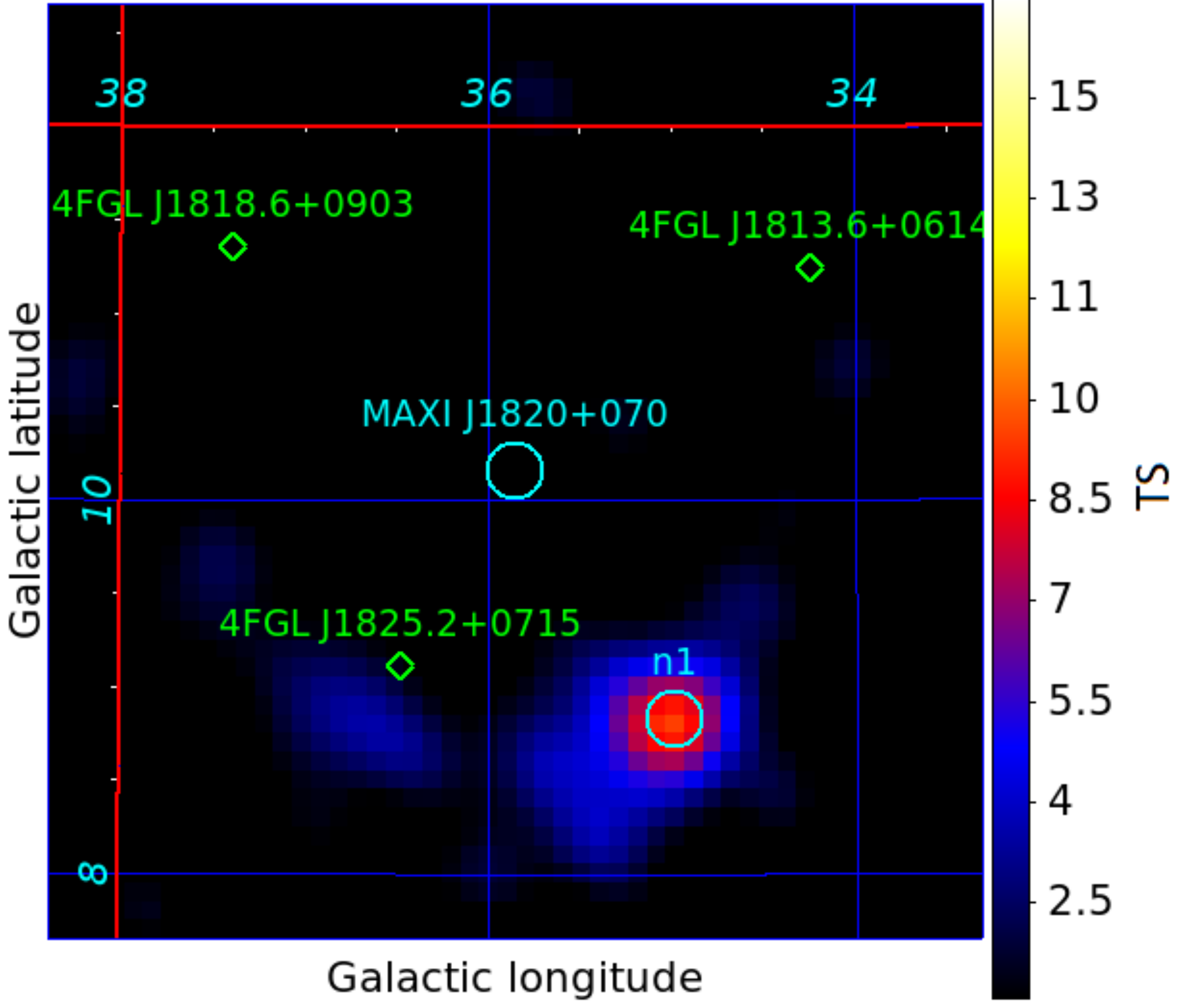}
    \quad
    \includegraphics[width=0.66\textwidth]{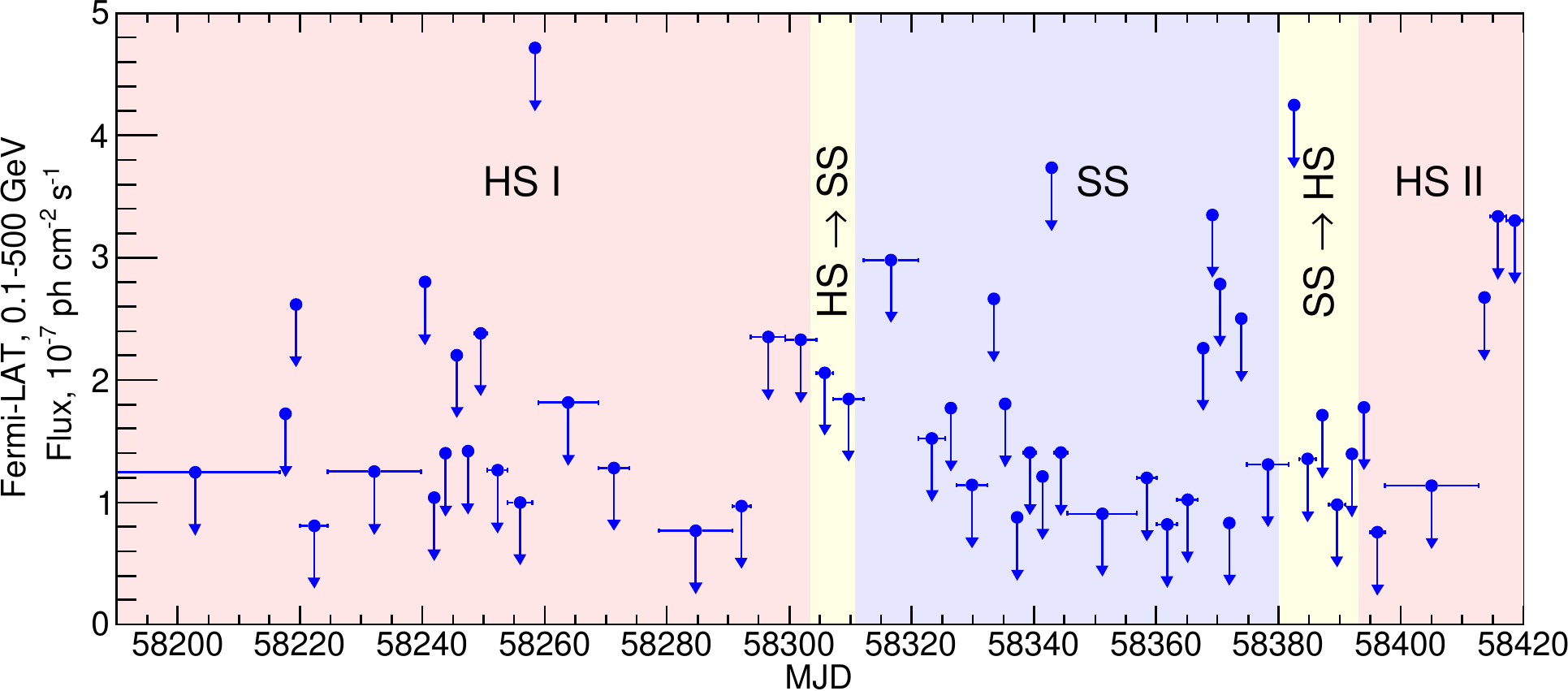}
    \caption{\textit{Left panel}: Test-statistics map in Galactic coordinates of a $5^\circ \times 5^\circ$ region around MAXI~J1820+070 (cyan circle at the centre) in the $0.1-500$~GeV energy range, with a $0.1^\circ$ pixel size. Green symbols show the 4FGL \textit{Fermi}-LAT sources present in the region. The cyan circle marked as \textit{n1} shows a TS $\sim 10$ point-like residual positionally coincident with PSR~J1823+0550.  \textit{Right panel}: \textit{Fermi}-LAT light curve of MAXI~J1820+070 in the $0.1-500$~GeV energy band with an adaptive time binning. The bin widths correspond to 16 photons arrived in a $1^\circ$-radius around MAXI~J1820+070. Source states are represented with light red (HS), light blue (SS) and light yellow (HS--SS / SS--HS) background colours.}
    \label{fig:ts_map_and_lc}
\end{figure*}

In order to check the quality of the considered model of the region at the initial step of the analysis, we built a test-statistics (TS) map showing the TS value of a point-like source not present in the model located in a given pixel of the map. The TS map obtained is shown in the left panel of Fig.~\ref{fig:ts_map_and_lc}. The map illustrates that the selected model describes the region well in the energy and time ranges considered. We note the presence of a TS $\sim 10$ residual at RA $= 275.88^\circ$, Dec $= 5.84^\circ$ (with a positional uncertainty of $0.15^\circ$), marked as \textit{n1} on the map. This residual is positionally coincident with PSR~J1823+0550 (PSR~B1821+05). We modelled this source as a point-like source with a power-law spectrum, ${\rm d}N/{\rm d}\varepsilon = K \varepsilon^{-\alpha}$, with $K$ being the flux normalisation. The best-fit parameters in the selected time range and in the $0.1-500$~GeV energy band are $K = (1.3 \pm 0.4) \times 10^{-12}$~ph cm$^{-2}$ s$^{-1}$ MeV$^{-1}$ at 1~GeV, and $\alpha = 2.3 \pm 0.2$. We included this source to the considered model of the region with a free normalisation and the index fixed to the best-fit value. After all these steps, MAXI~J1820+070 was not detected with the binned likelihood analysis for the time period considered (assuming a power-law spectrum model with a free spectral index), and is therefore not present at the TS map of Fig.~\ref{fig:ts_map_and_lc}.

In what follows, the \textit{Fermi}-LAT flux upper limits for MAXI~J1820+070 were calculated at a $0.95$~C.L. with the help of the \texttt{IntegralUpperLimit} module provided as a part of standard \textit{Fermi}-LAT data analysis software for a power-law index fixed to $\alpha = 2.5$, as for the VHE data analysis \citep[and also similar to what is observed for high-mass microquasars in the \textit{Fermi}-LAT energy range, see, e.g.,][]{zanin16,zdziarski18}. 

In order to search for a possible short-timescale variability observed in several microquasars detected up to GeV energies, we computed the light curve of the source with variable (adaptive) time binning \citep[e.g.,][]{lott12}. Namely, we selected a time bin duration such that each bin receives 16 photons in the $0.1-500$~GeV energy range and in a radius of $1^\circ$ around the MAXI~J1820+070 position. This resulted in 55 time bins in the total time range considered with an average duration of $4.2$~days (minimum: 1~day, maximum: 27~days). Such time bin selection allows us to identify the shortest possible periods during which the source potentially could be detected with up to a $\sim 4\sigma$ significance. A similar approach was found to be effective for a search of short flares during the periods of strong GeV variability of PSR~B1259-63 in the analysis of \cite{psrb20,psrb21}. The performed analysis did not result in the detection of MAXI~J1820+070 in any of the time bins, and the resulting ULs are shown in the right panel of Fig.~\ref{fig:ts_map_and_lc}.

\begin{table}
    \centering
    \begin{tabular}{c c}
    \hline
    \hline
    Parameter                   & Value                             \\
    \hline
    Response functions          & P8R3\_V3                          \\
    evclass                     & $128$                             \\
    evtype                      & $3$                               \\
    zmax                        & $90^\circ$                        \\
    Spatial bin width           & $0.05^\circ$                      \\
    Energy bins per decade      & $5$                               \\
    edisp\_bins                 & $-3$                              \\
    Energy range                & 0.1 -- 500~GeV                    \\
    ROI (free)                  & $14^\circ$                        \\
    ROI (fixed)                 & $14^\circ$ -- $24^\circ$          \\
    Catalogue                   & 4FGL-DR3                          \\
    Background (galactic)       & gll\_iem\_v07.fits                \\
    Background (isotropic)      & iso\_P8R3\_SOURCE\_V3\_v1.txt     \\
    Likelihood analysis optimiser& NEWMINUIT \\
    Time ranges                 & 542505605 -- 562464005 (TOTAL)    \\
                                & 542505605 -- 552398405 (HS I)     \\
                                & 552398405 -- 553020485 (HS--SS)   \\
                                & 553020485 -- 559008005 (SS)       \\
                                & 559008005 -- 560131205 (SS--HS)   \\
                                & 560131205 -- 562464005 (HS II)    \\
    \hline
    \end{tabular}
    \caption{Details of \textit{Fermi}-LAT data analysis. From top to bottom, the parameters of the analysis are: type of response functions; event class and type; maximum zenith angle; spatial and energy bins widths for the likelihood analysis; number of energy dispersion bins; energy range used for the analysis; radius of the region of interest up to which the sources were included with free normalisation (ROI (free)); radii range in which the sources were included with all parameters fixed to 4FGL-DR3 values (ROI(fixed)); used catalogue; Galactic diffuse and isotropic diffuse background templates; time range (in \textit{Fermi} seconds) used for the analysis.}
    \label{tab:fermi_analysis_details}
\end{table}

\subsection{Additional multiwavelength data}\label{sec:MWL}

Data from several radio telescopes at different frequencies are taken from \cite{bright20}. Optical data are taken from \cite{celma19}, in which observations performed with the Joan Or\'o Telescope \citep[TJO;][]{colome10} and \textit{Swift} Ultraviolet/Optical Telescope \citep[\textit{Swift}/UVOT;][]{swift05} are reported. The optical fluxes are obtained from images taken with the 5 Johnson-Cousins filters (with central wavelengths around 366, 435, 548, 635 and 880~nm, respectively from the $U$ to $I$ filters), and they are already corrected for interstellar extinction with values ranging from 0.2 to 0.6 mag \citep[see][and references therein]{fitzpatrick99,celma19}. Public LCs from MAXI/GSC \citep{maxi09} and \textit{Swift} Burst Alert Telescope \citep[\textit{Swift}/BAT][]{Krimm_2013} are also included.

For the spectral energy distributions (SEDs) of MAXI~J1820+070, International Gamma-Ray Astrophysics Laboratory \citep[INTEGRAL][]{integral03} data are added to that of the previously mentioned instruments, based on results by \cite{roques19} from MJD~58206 to 58246, during the first half of the HS. Data from the Neutron star Interior Composition Explorer \citep[NICER,][]{NICER} are also used. NICER is designed to study neutron stars via soft X-ray timing spectroscopy and has been operating from the International Space Station since 2017. It observed for 109~h, 21.8~h and 4.56~h during the HS, the HS--SS transition and SS--HS transition, respectively. Pre-processed event files were retrieved through the HEASARC database. Re-processing and filtering were done using standard criteria with the \texttt{nicerl2} task from the NICERDAS software available in the HEAsoft distribution\footnote{\url{https://heasarc.gsfc.nasa.gov/docs/software/lheasoft/}} (v6.26). Spectra were extracted using the \textit{extractor} function from the ftools package. Error bars account only for the statistical uncertainty on detector counts, namely ±1 standard deviation of a Poisson distribution. Energy and gain calibrations were performed using the HEASARC Calibration Database version XTI(20200722). To avoid telemetry saturation, the fraction of active modules had to be adjusted. This was taken into account considering that each module contributes equally to the effective area. The fluxes were corrected for interstellar extinction using a hydrogen column density of $N_{\rm H} = 1.4\times10^{21}$~cm$^{-2}$ \citep{dzielak21}. 


\section{Results}\label{sec:results}

The observations of MAXI~J1820+070 reported in this work do not show any significant emission in either HE or VHE gamma rays, regardless of the source state. The computed integral flux ULs for \textit{Fermi}-LAT data and the combined dataset of H.E.S.S., MAGIC and VERITAS are shown in Table~\ref{tab:intUL} for each X-ray state. The former are calculated for photon energies $\varepsilon > 100$~MeV, while the latter are computed at $\varepsilon > 200$~GeV for the HS, the HS--SS transition, the SS and the whole sample, and at $\varepsilon > 300$~GeV for the SS--HS transition. The increase in energy threshold of the last data set is due to a higher average zenith angle of the observations, which does not allow electromagnetic showers triggered by lower energy gamma rays to be detected. We note that the VHE UL for the SS, shown here for completeness, may not be representative of the whole source state, since it only covers the very first days of the SS (see the top panel of Fig.~\ref{fig:HR}).

\begin{figure*}
	\centering
    \includegraphics[width=0.49\linewidth]{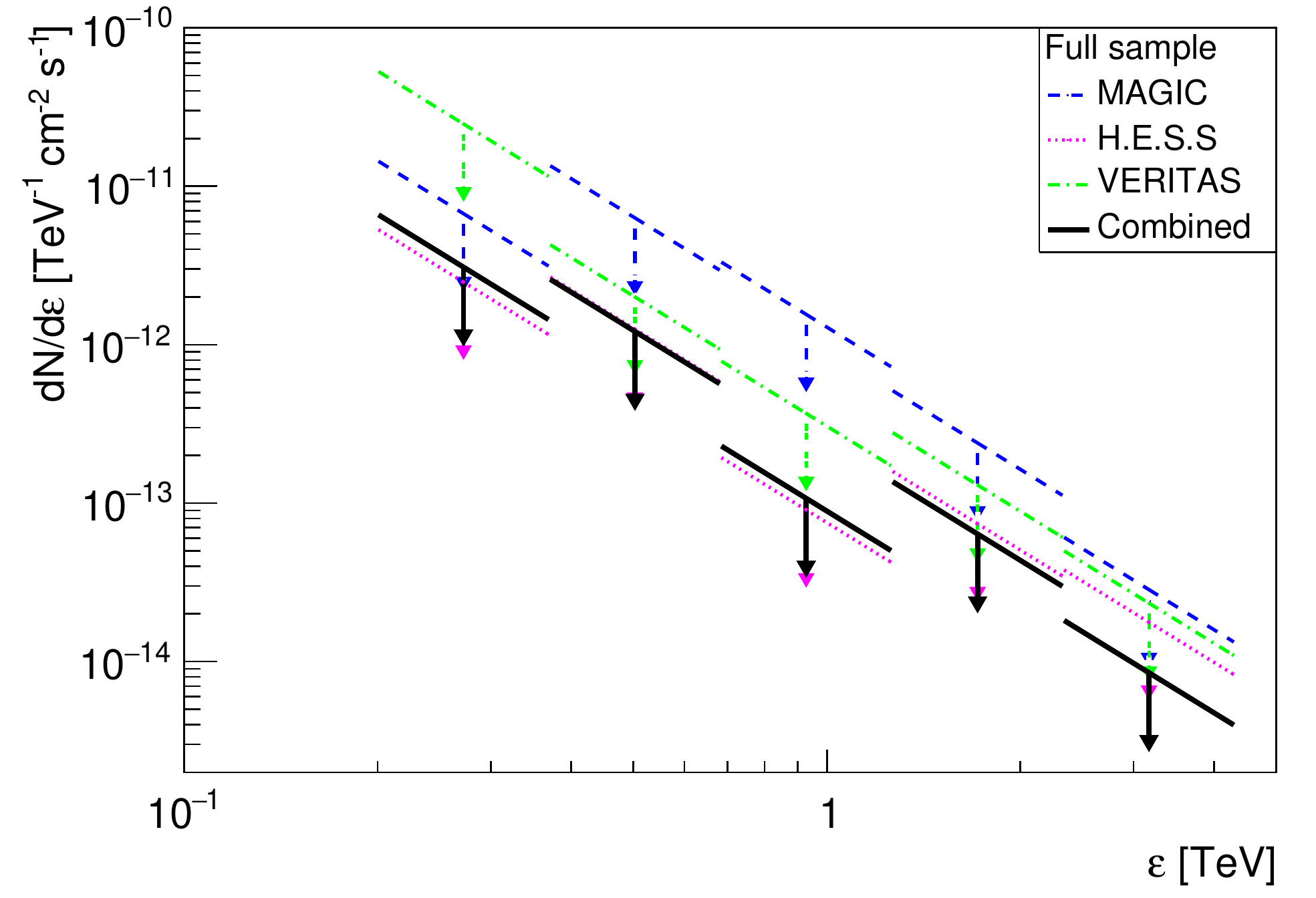}
    \includegraphics[width=0.49\linewidth]{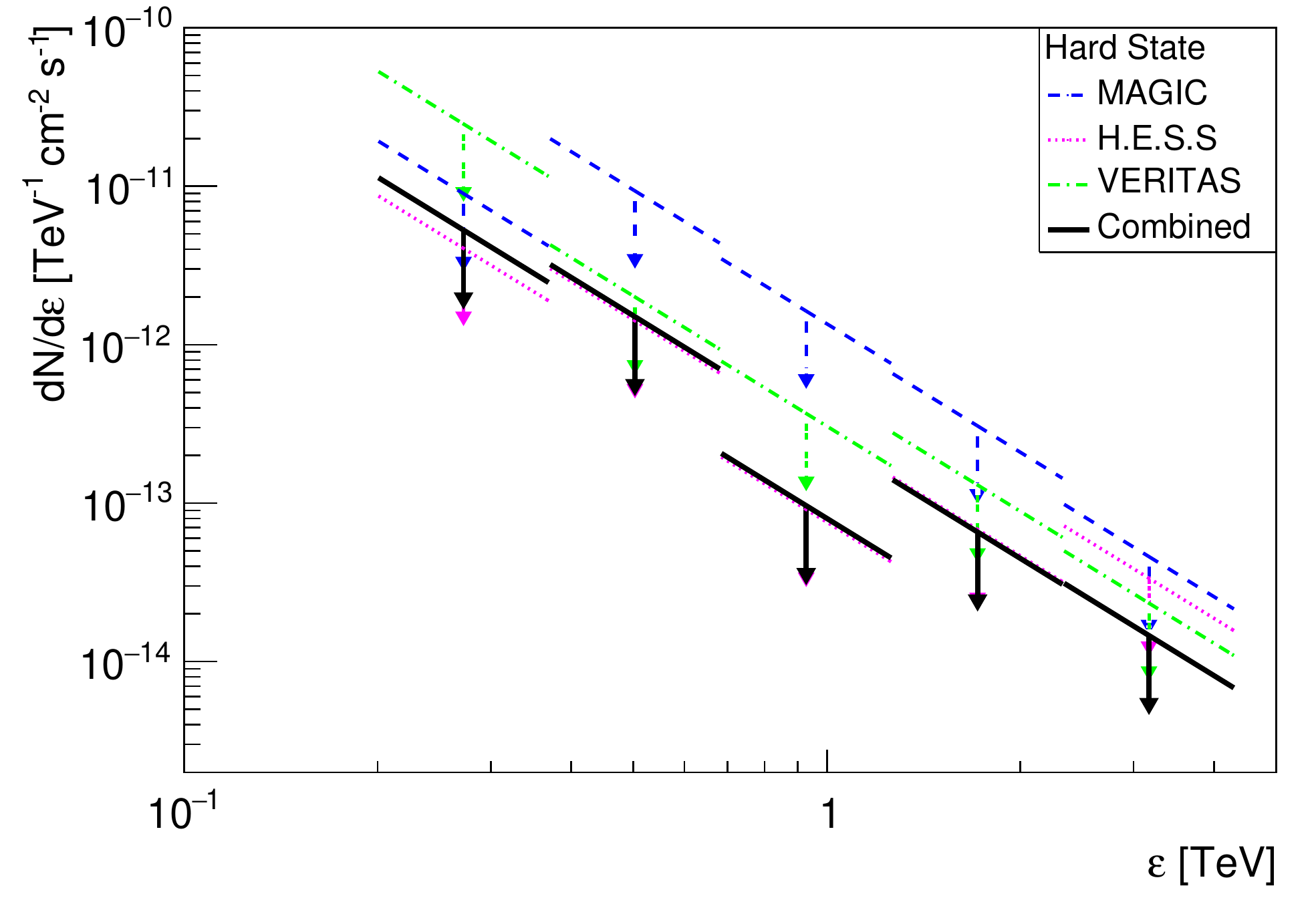}
    \vskip 3mm
    \includegraphics[width=0.49\linewidth]{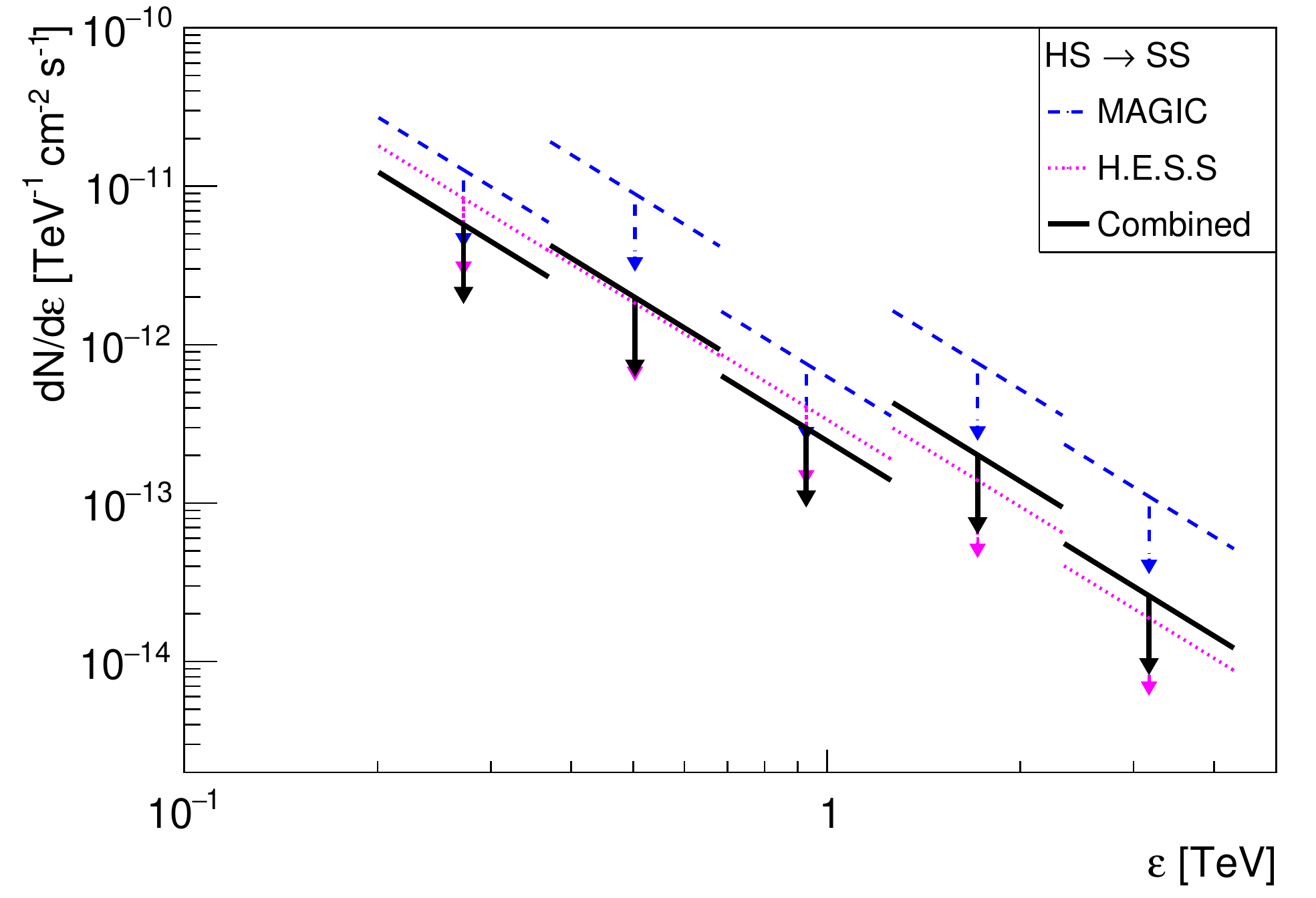}
    \includegraphics[width=0.49\linewidth]{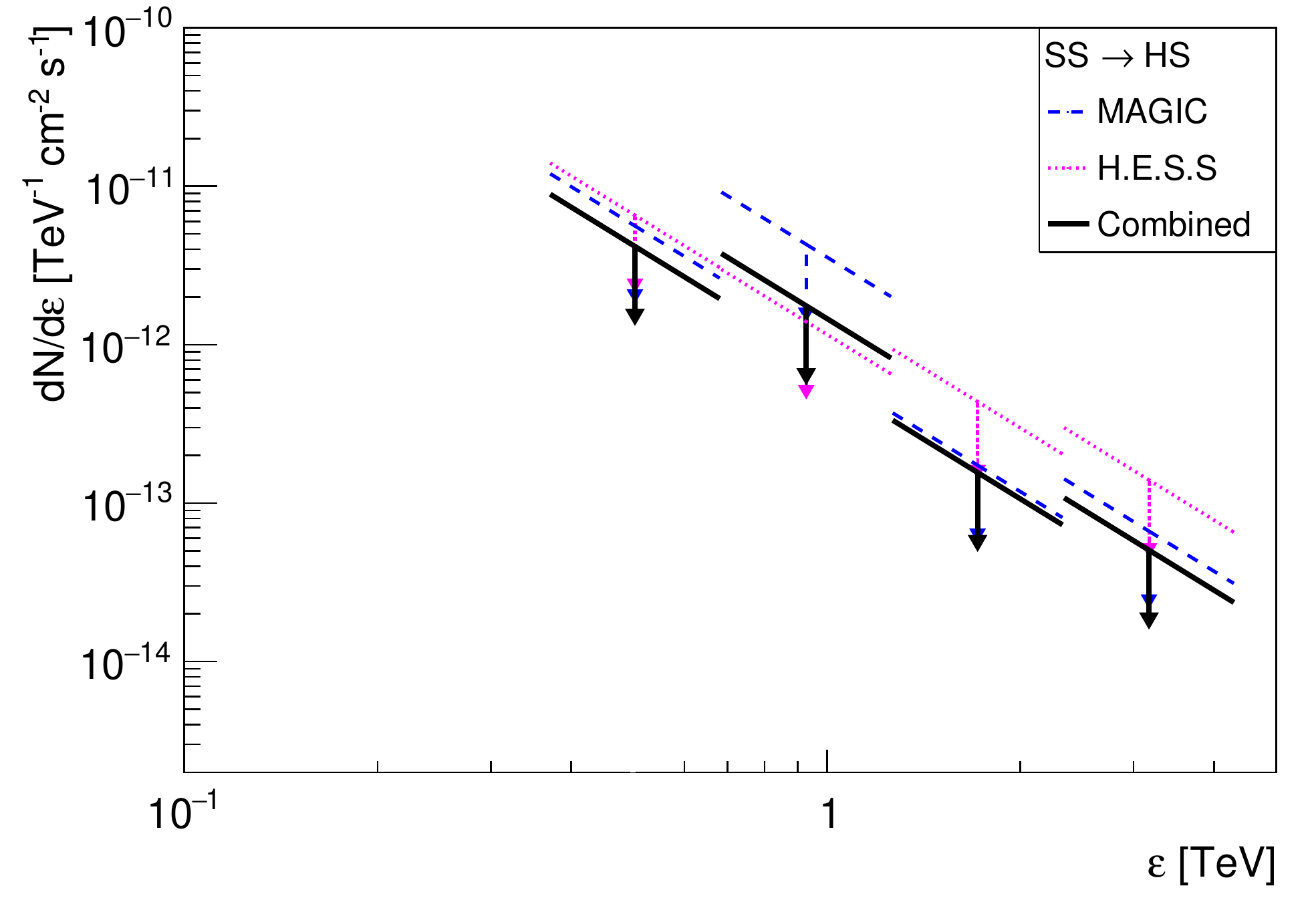}
    \caption{Differential flux upper limits of MAXI~J1820+070 for different energy bins and source states (indicated in the legend). Coloured markers represent the results for the individual experiments, while black lines show the combined upper limits.}
    \label{fig:diffUL}
\end{figure*}

Fig.~\ref{fig:diffUL} presents the ULs on the VHE differential flux obtained for 5 different energy bins and each source state for which observations were performed (excluding the poorly-covered SS). Both individual and combined ULs are shown in each case. For the SS--HS transition, the lowest energy bin is not computed due to the increased energy threshold of the corresponding observations. The differences between individual ULs in the same energy bin originate from the different instrument sensitivities and observation times, as well as from statistical fluctuations (see Appendix~\ref{sec:fluxComputation}). Except in the case of significant differences between the individual ULs, the combined ULs are tighter than any of the individual ones.


\begin{table}
    \begin{center}
    \caption{Integral flux upper limits with a 0.95 C.L. during different source states, between 0.1 and 500~GeV from \textit{Fermi}-LAT data, and above 200~GeV from the combined H.E.S.S., MAGIC and VERITAS data. For the SS $\rightarrow$ HS transition, the UL above 300~GeV is shown instead.}
    \begin{tabular}{l c c}
    \hline \hline
                            & \textit{Fermi}-LAT UL     & IACT UL                   \\
    Source state            & ($0.1-500$~GeV)           & ($>200/300$~GeV)          \\
                            & [ph cm$^{-2}$ s$^{-1}$]   & [ph cm$^{-2}$ s$^{-1}$]   \\
    \hline
    Hard State I            & $3.1\times10^{-8}$        & $9.5\times10^{-13}$       \\
    HS $\rightarrow$ SS     & $1.6\times10^{-7}$        & $9.5\times10^{-13}$       \\
    Soft State              & $2.5\times10^{-8}$        & $1.6\times10^{-12}$       \\
    SS $\rightarrow$ HS     & $5.2\times10^{-8}$        & $2.2\times10^{-12}$       \\
    Hard State II           & $6.0\times10^{-8}$        & $-$                       \\
    TOTAL                   & $1.8\times10^{-8}$        & $7.2\times10^{-13}$       \\
    \hline
    \end{tabular}
    \label{tab:intUL}
    \end{center}
\end{table}

\begin{figure}
	\centering
    \includegraphics[width=\linewidth]{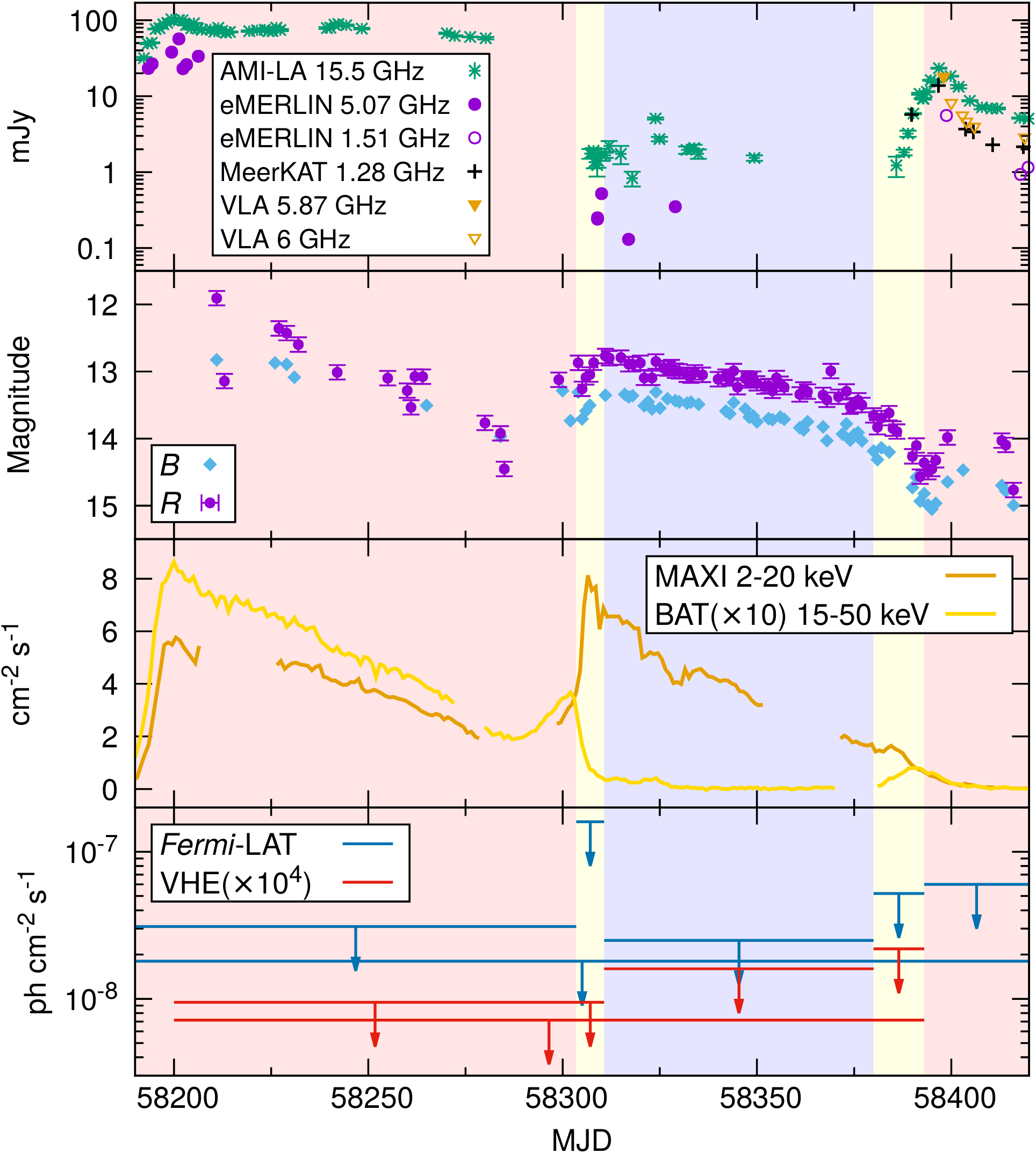}
    \caption{From \textit{top} to \textit{bottom}: Radio, optical, X-ray, and gamma-ray light curves of MAXI~J1820+070 during its 2018 outburst (MJD~58189.0 -- 58420.0). The shaded areas correspond to the HS (light red), the HS--SS and SS--HS transitions (light yellow), and the SS (light blue). The units of the third panel are [ph cm$^{-2}$ s$^{-1}$] for MAXI/GSC, and [counts cm$^{-2}$ s$^{-1}$] for \textit{Swift}/BAT. The latter fluxes are multiplied by 10 for a better visualisation. The bottom panel shows the \textit{Fermi}-LAT ULs above 100~MeV, and the H.E.S.S./MAGIC/VERITAS combined ULs (multiplied by $10^4$) for each source state and transition for which data are available, as well as for the whole outburst. The VHE ULs are computed above 200~GeV except for the SS--HS transition (MJD~58380.0 -- 58393.0) for which 300~GeV ULs are shown.}
    \label{fig:LC}
\end{figure}

Fig.~\ref{fig:LC} shows the LCs of MAXI~J1820+070 at different frequencies, with the gamma-ray LC corresponding to the ULs in Table~\ref{tab:intUL}. The radio fluxes in the top panel include both the core emission from the jet regions close to the binary system, and the radiation emitted by discrete ejections launched during the HS--SS transition. Core emission is dominant during the source HS, while the ejections dominate throughout the SS, during which no core emission is detected \citep[see][for the details]{bright20}. The optical fluxes in the second panel are obtained from a total set of 16457 images distributed over 113 different nights between March and November 2018 \citep{celma19}. The X-ray LCs in the third panel are obtained from the daily fluxes of MAXI~J1820+070 from MAXI/GSC (for 2~keV $\leq \varepsilon \leq 20$~keV) and \textit{Swift}-BAT (for 15~keV $\leq \varepsilon \leq 50$~keV). The gaps represent the periods when the source was not observed with these instruments.

\begin{figure}
	\centering
    \includegraphics[width=\linewidth]{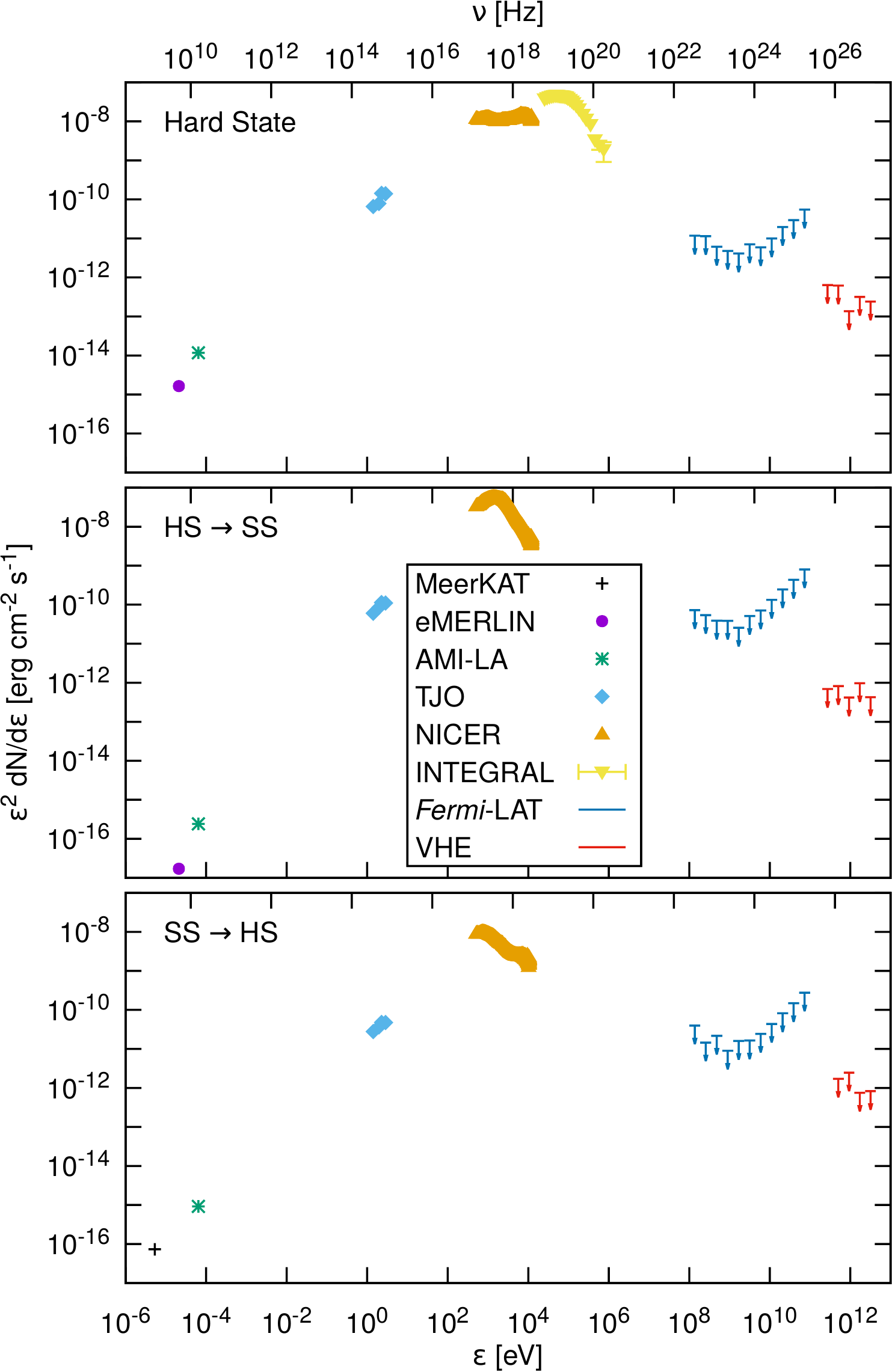}
    \caption{From \textit{top} to \textit{bottom}: Spectral energy distributions of MAXI~J1820+070 averaged over the HS, the HS--SS transition, and the SS--HS transition. \textit{Fermi}-LAT points include the contribution from the two HSs of the source. The eMERLIN data shown are those at 5.07~GHz. MeerKAT data are used in the bottom panel instead of eMERLINs.}
    \label{fig:SEDs}
\end{figure}

The SEDs of MAXI~J1820+070, averaged for those source states well represented by the VHE data, are shown in Fig.~\ref{fig:SEDs}. We note that the jump between NICER and INTEGRAL data in the top panel is just an effect of the different time coverage of the observations. While NICER data are averaged over the whole duration of the HS, INTEGRAL data only cover roughly the first half of it (see Sect.~\ref{sec:MWL}), when the average X-ray flux was higher.

\section{Discussion}\label{sec:discussion}

In this section, we provide a short description of multiwavelength measurements of MAXI~J1820+070 from radio to X-rays. Based on some assumptions regarding the extrapolation of the jet synchrotron spectrum and the distribution (energy dependence and maximum energy) of the non-thermal particles, we estimate the expected jet emission in HE and VHE gamma rays analytically. This expected emission is then compared to the measured ULs, and used to constrain the properties of a potential gamma-ray emission region in MAXI~J1820+070.

\subsection{Multiwavelength overview of the source}

Radio emission from MAXI~J1820+070 provides evidence for jet activity during the whole 2018 outburst. This emission is dominated by a steady jet in the HS, a discrete blob in the HS--SS transition (the emission of which is also dominant throughout the SS as the blob moves away from the binary system), and a jet rebrightening during the SS--HS transition \citep{bright20}. Without accounting for blob emission during the SS, the radio and hard X-ray fluxes have similar behaviours: they decrease slowly through the HS, have a steep decrease in the HS--SS transition, are practically undetectable during the SS, and increase again in the SS--HS transition. This is expected from the standard picture of BH-LMXBs, where steady radio jets in the HS coexist with a hard X-ray emitting corona, both of them disappearing in the SS. During the HS, synchrotron emission from the jet is likely the dominant contribution to the SED up to infrared frequencies, beyond which the spectrum becomes dominated by disk and coronal emission \citep{rodi21,tetarenko21,zdziarski21_pair}. We nonetheless note that the jets in LMXBs may still contribute significantly up to hard X-rays through their synchrotron emission \citep[e.g.][]{markoff05}.

Regarding the contribution from the star to the overall SED, MAXI~J1820+070 had a magnitude of 17.4 in the $G$ filter -- with a central wavelength around 460~nm -- before the outburst \citep{simbad00,gaia18}, which is at least 3 magnitudes above the $B$ magnitude during the flare (from the filters shown in the second panel of Fig.~\ref{fig:LC}, the $B$ filter is the closest one to the $G$ filter). This means that the optical flux of MAXI~J1820+070 during the flare was at least about 15 times larger than before the outburst. Nonetheless, this increase in flux cannot be exclusively associated to the brightening of the accretion disk, since the stellar luminosity can also increase during the outburst owing to the heating of the stellar surface produced by the X-ray emission close to the BH \citep[e.g.][]{dejong96}. Assuming a stellar radius of $\sim 10^{11}$~cm equal to that of the Roche lobe, the solid angle of the star as seen by the BH is $\sim 0.01$~sr. This means that the optical luminosity of the X-rays reprocessed in the stellar surface should be about 2 orders of magnitude below the X-ray luminosity. This optical luminosity is comparable to what is observed from the whole system (see Fig.~\ref{fig:SEDs}), so we can conclude that the stellar contribution to the optical flux of MAXI~J1820+070 may be significant.

\subsection{Analytical estimates}

For the estimates performed in this section, the particles responsible for the non-thermal emission are assumed to be only electrons (and positrons), although we note that hadrons might contribute to the overall emission of the system \citep[see, e.g.,][for typical electron and proton cooling timescales in microquasar environments]{bosch09b}. These electrons are likely accelerated up to relativistic energies close to the BH, and different acceleration mechanisms may play a role \citep[see, e.g.,][for a description of different processes that can contribute to particle acceleration]{bosch12}. The non-thermal emission of the electrons is assumed to come from their synchrotron and IC cooling. The derived jet inclination and speed in MAXI~J1820+070 make the counter-jet emission significantly more deboosted than that from the jet. The discussion can therefore be focused on the jet emission, and the counter-jet contribution can be neglected. In the following, primed quantities refer to the reference frame moving with the jet flow, while unprimed ones refer to quantities in the laboratory frame or as seen by the observer.

\subsubsection{A steady jet in the hard state}\label{sec:disc_hs}

Some jet properties during the initial HS of the source were constrained by \cite{zdziarski22_jet} based on the radio to optical emission. In particular, the synchrotron break frequency (above which the emission becomes optically thin) is measured to be $\nu_0 \approx 2\times10^4$~GHz. They also find that for a jet Lorentz factor of $\Gamma \approx 3$, the onset of the jet synchrotron emission comes from a distance to the BH of $r \approx 3.8\times10^{10}$~cm, where the magnetic field is $B' \approx 10^4$~G if equipartition between the magnetic and particle energy densities is assumed \citep[the equipartition condition approximately corresponds to the minimum energy requirement for synchrotron radiation, which happens for a magnetic energy of $\sim 0.75$ times the particle one; e.g.,][]{longair81}.

To estimate the gamma-ray emission of the source, we assume that gamma rays are produced by IC scattering of photons coming from the accretion disk or the corona by jet electrons. This means that the target photons reach the jet mainly from behind, which is very likely the case for X-rays in MAXI~J1820+070, and is also approximately the case for optical photons. Given the conditions in the source, the estimates can be done in the context of the Thompson regime, which is approximately valid at the adopted energies ($\gamma \varepsilon \lesssim m_ec^2$, see below), and simplifies the calculations \citep[e.g.,][]{longair81}. In this regime, IC is more efficient and the energy gain of the scattered photons is proportional to the square of the electron Lorentz factor $\gamma$. 

For HE gamma rays, a characteristic energy of $\varepsilon = 100$~MeV is taken, which would be the result of the IC scattering towards the observer of target X-ray photons with typical energies of $\sim 1$~keV by $E'_{\rm HE} \sim 250$~MeV electrons ($\gamma' \sim 500$). These are reference values for which data at the target photon energy are available, although target photons with energies similar to the chosen ones would also contribute to the IC emission around $\varepsilon = 100$~MeV. The electrons with energy $E'_{\rm HE}$ emit synchrotron photons with an observed frequency of $\nu_{\rm syn} \approx 1.5 \times 10^6$~GHz. The observed flux density at this frequency \citep[extrapolated from the infrared data in][]{zdziarski22_jet} is $F_\nu^{\rm syn} \sim 30$~mJy. The observed IC flux of the electrons with energy $E'_{\rm HE}$ can then be estimated as:
    \begin{linenomath}\begin{equation}\label{flux}
    \varepsilon^2 \left. \! \frac{{\rm d}N}{{\rm d}\varepsilon}\right| _{\rm IC} = \nu_{\rm IC} F_\nu^{\rm IC} \approx \nu_{\rm syn} F_\nu^{\rm syn} \frac{\dot{E}'_{\rm IC}}{\dot{E}'_{\rm syn}} \ ,
    \end{equation}\end{linenomath}
where $\dot{E}'_{\rm IC} = - 0.039 \, u' E'^2$ and $\dot{E}'_{\rm syn} = - 1.6 \times 10^{-3} B'^2 E'^2$ are the IC and synchrotron cooling rates in cgs units, respectively. The energy density of the target photon field with luminosity $L_{\rm tar} \sim 2\times10^{37}$~erg~s$^{-1}$ is $u'\approx u/\Gamma^2 = L_{\rm tar} / 4\pi r^2 c \Gamma^2$ \citep[valid as long as the target photons reach the jet from behind;][]{dermer94}. Equation~\eqref{flux} yields a predicted IC energy flux at $100$~MeV of $\sim 6 \times 10^{-13}$~erg~cm$^{-2}$~s$^{-1}$, about a factor of 20 smaller than the obtained ULs at this energy (see upper panel of Fig.~\ref{fig:SEDs}). We note that the predicted energy flux increases with the ratio $\eta$ of particle-to-magnetic energy density as $\nu_{\rm IC} F_\nu^{\rm IC} \propto \eta^{0.35}$, as long as the corresponding $\nu_{\rm syn}$ is in the optically thin regime \citep[see the dependency of $B'$ with the energy density fraction in][and how this changes the values of $\nu_{\rm syn}$ and $F_\nu^{\rm syn}$]{zdziarski22_jet}. For example, taking a value of $\eta = 100$ raises the expected energy flux to $\sim 3 \times 10^{-12}$~erg~cm$^{-2}$~s$^{-1}$, a factor of $5$ higher than in equipartition and only $4$ times smaller than the ULs. 

Regarding the VHE emission, an extrapolated power-law electron distribution is assumed, i.e. $\mathcal{N}'(E') \propto E'^{-p}$, with $\mathcal{N}'(E')$ being the number of electrons per energy unit. This distribution is taken up to $E'_{\rm VHE} \sim 500$~GeV, which is the energy required to emit VHE gamma rays with $\varepsilon \sim 200$~GeV through IC with optical target photons. In order not to contradict the observations, a soft injection index of $p \gtrsim 3$ is required for the high-energy electrons (with energies above those responsible for the infrared emission reported in \citealt{zdziarski22_jet}). Otherwise, the observed MeV fluxes would be violated by the synchrotron emission of the electrons with hundreds of GeV, and the VHE ULs would be violated by the expected IC emission of these electrons. On the other hand, using $p \gtrsim 3$ yields an expected VHE emission that falls at least 2 orders of magnitude below the obtained UL for the lowest VHE bin in Fig.~\ref{fig:SEDs}. Therefore, the obtained VHE ULs are not so constraining as the HE ones in the source HS.

\subsubsection{Discrete ejections during the state transitions}\label{sec:disc_transitions}

For the HS--SS transition, \cite{bright20} determined that the radio emission was dominated by a discrete blob of plasma. The estimates in this section assume, for both state transitions\footnote{We note that the blob model may not hold for the SS--HS transition, although it is still used for the sake of simplicity and due to the lack of much information for this source state.}, a one-zone spherical radio emitter in the flow frame with the Lorentz factor and inclination values reported in Sect.~\ref{sec:introduction}. The non-thermal electrons responsible for this synchrotron radio emission are taken as reference to obtain the expected IC emission in the source. We use the spectral shape derived from the two radio points in Fig.~\ref{fig:SEDs}, which indicates a self-absorbed synchrotron emission at the observed frequencies. Therefore, the break frequency should be located at a frequency higher than $15.5$~GHz. On the other hand, for the HS--SS transition this frequency has to be lower than $\sim 700$~GHz if $p\sim 2$, since otherwise the optical fluxes would be violated by the blob synchrotron emission. For simplicity, a break frequency of $\nu_0 = 100$~GHz is used in both state transitions: if $\nu_0$ were lower, the limits derived below would be less restrictive, and more restrictive otherwise.With the Doppler boosting factor of the blob being $\delta = [\Gamma (1-\beta \cos\theta)]^{-1}$, the break frequency value in the flow frame is $\nu'_0 = \nu_0/\delta$, and the extrapolated flux density at $\nu'_0$ is $F'_0 = F_0/\delta^3$. The distance to the source $d$ can be used to constrain the magnetic field $B'$ and radius $R'$ of the radio-emitting blob through the following relation in cgs units \citep[derived from Eq. 6.38 in][]{pacholczyk70}:
    \begin{linenomath}\begin{equation}\label{nuSSA}
    \nu'_0 \approx 10^{12} F_0^{\prime \, 2/5} B^{\prime \, 1/5} R'^{-4/5} d^{4/5} \ .
    \end{equation}\end{linenomath}
We note that, due to the high inclination of the system, $R'$ is approximately equal to $R$ measured in the direction of the observer. The magnetic field is parametrized through the fraction $\eta$ of magnetic to particle energy density:
    \begin{linenomath}\begin{equation}\label{B}
    \frac{B^{\prime 2}}{8\pi} = \eta \frac{E'_{\rm NT}}{V'} \ ,
    \end{equation}\end{linenomath}
where $V' = 4\pi R'^3 / 3$ is the proper volume of the emitting region, and $E'_{\rm NT}$ is the energy budget of the non-thermal electrons in the blob. Taking an electron distribution $\mathcal{N}'(E') = Q E'^{-p}$ with $p = 2$ between $E'_{\rm min} \sim 50$~MeV and $E'_{\rm max} \sim$~GeV yields
    \begin{linenomath}\begin{equation}\label{ENT}
    E'_{\rm NT}=Q \ln(E'_{\rm max} / E'_{\rm min}) \sim 3Q \ ,
    \end{equation}\end{linenomath}
with $Q$ being a normalisation constant. The values of $p$, $E'_{\rm min}$ and $E'_{\rm max}$ are not strongly constrained by the observations, but Eq.~\eqref{ENT} is not very sensitive to the exact values of $E'_{\rm min}$ and $E'_{\rm max}$ as long as $p \sim 2$. Additionally, the optically thin synchrotron spectral luminosity can be related to the particle distribution as:
    \begin{linenomath}\begin{equation}\label{Le}
    L'_{\varepsilon'} \approx \mathcal{N}'(E') |\dot{E}'_{\rm syn}| \frac{{\rm d}E'}{{\rm d}\varepsilon'} \ .
    \end{equation}\end{linenomath}

Taking an equipartition magnetic field with $\eta = 1$, Eqs.~\eqref{nuSSA} to \eqref{Le} provide the following results for the radio emitter in the HS--SS (SS--HS) transition: a radius of $R' \approx 2.0 \times 10^{11}$~cm ($1.1 \times 10^{11}$~cm), a non-thermal energy budget of $E'_{\rm NT} \approx 2.8 \times 10^{36}$~erg ($6.5 \times 10^{35}$~erg), and a magnetic field of $B' \approx 47$~G ($55$~G).

The same reference energy as in the HS is taken for the HE emission through IC, i.e. $\varepsilon = 100$~MeV, which results from the scattering of $\sim 1$~keV photons by $E'_{\rm HE} \sim 200$~MeV electrons. For the VHE emission, a characteristic energy of $\varepsilon = 200$~GeV ($400$~GeV) is chosen for the HS--SS (SS--HS) transition. These gamma-ray photons are the result of the scattering of disk target photons with typical energies of $\sim 0.5$~eV by $E'_{\rm VHE} \sim 400$~GeV ($600$~GeV) electrons. Observational evidence of non-thermal electrons is available only for energies below 10~MeV in the flow frame, since the electrons in this energy range are responsible for the emission up to $15.5$~GHz. Therefore, an extension of the electron distribution up to the required $E'_{\rm HE}$ and $E'_{\rm VHE}$ is assumed with a power-law shape with index $p = 2$. Reaching such particle energies is not unreasonable for the equipartition magnetic field obtained above, since high acceleration efficiencies are not required to reach those energies. The most constraining efficiency needed is $\eta_{\rm acc} \lesssim 200$ for $E' = 600$~GeV, with $t'_{\rm acc} = \eta_{\rm acc} E'/e c B'$ being the acceleration timescale.

To estimate the expected IC emission, a similar method to the one developed for the HS is used, following Eq.~\eqref{flux}, where $\nu_{\rm syn}$ and $F_\nu^{\rm syn}$ are now the characteristic synchrotron frequency of the electrons with each of the aforementioned energies, $E'_{\rm HE}$ and $E'_{\rm VHE}$, and the corresponding extrapolated flux at $\nu_{\rm syn}$, respectively. This sets a minimum distance of the potential gamma-ray emitter to the BH through the dependence of the target photon energy density $u'$ with this distance. If the emitter were closer to the BH (and the accretion disk), $u'$ would be so high that the gamma-ray ULs would be violated by the IC emission of the blob. For the HS--SS (SS--HS) transition, the derived distances of the potential HE and VHE emitters to the BH are $r_{\rm HE} \gtrsim 2.8 \times 10^{10}$~cm ($1.0 \times 10^{10}$~cm) and $r_{\rm VHE} \gtrsim 1.1 \times 10^{12}$~cm ($2.3 \times 10^{11}$~cm). The radio emitter size constraints impose that $r_{\rm HE}$ is likely larger, at least a few times $10^{11}$~cm. Conversely, both $r_{\rm HE}$ and $r_{\rm VHE}$ cannot be more than several times the radio emitter size, the exact value depending on the blob expansion velocity, since that would increase the emitter size to the point that the synchrotron radio emission would become optically thin below $\nu_0$.

Both the potential HE and VHE emitter distances derived above are sufficiently large for the blob gamma-ray emission to be unaffected by gamma-gamma absorption with the external photons from the disk and the corona (this also applies to the HS emitter studied in Sect.~\ref{sec:disc_hs}). Also, for the values of $R'$ and $B'$ obtained, the energy density of the synchrotron soft X-ray photons (emitted as long as the electrons reach high-enough energies) should be comparable to that of the X-rays coming from the corona and/or the disk. Moreover, the energy density of the synchrotron optical photons should also be comparable to that of the disk optical photons. Therefore, synchrotron self-Compton should be responsible for a significant fraction of any HE and VHE emission of MAXI~J1820+070.


\section{Summary and conclusions}\label{sec:summary}

Observations of the exceptionally bright X-ray source MAXI~J1820+070 have been performed with the H.E.S.S., MAGIC and VERITAS experiments in VHE gamma rays. These data complement \textit{Fermi}-LAT observations of HE gamma rays, as well as additional multiwavelength data from radio to X-rays. The latter show the expected behaviour for a typical BH-LMXB, indicating a source following the usual HS--SS--HS cycle during an outburst.

Radio emission throughout the whole outburst provides evidence for the presence of jets with a population of non-thermal particles (electrons and possibly positrons) emitting via the synchrotron mechanism. Based on the study performed by \cite{zdziarski22_jet}, the estimated HE emission during the HS could be not far below the obtained ULs if the magnetic field is well below equipartition. Additionally, for a spherical blob-like radio emitter during the state transitions, significant constraints to a potential gamma-ray emitter in the source can be set using reasonable assumptions for the synchrotron transition frequency and the spectrum of the emitting electrons. For an equipartition magnetic field, the potential HE emitter should be located at a distance from the BH between a few $10^{11}$ and a few $10^{12}$~cm. If electrons are efficiently accelerated up to energies of $\sim 500$~GeV in the flow frame, a putative VHE emitter should also be located in a similar region, between $\sim 10^{11}$~cm and $10^{13}$~cm. Having the emitter closer than the region defined by these limits would violate the gamma-ray observations. Conversely, if the emitter is farther than this region, its emission would not be consistent with the observed optically-thick radio spectrum. The relatively narrow range of allowed distances during the transitions, and the inferred gamma-ray fluxes in the HS, indicate that the HE and VHE gamma-ray flux of MAXI~J1820+070 (and possibly other BH-LMXBs showing evidence for non-thermal emission) might not be too far from being detectable with the current instrument sensitivities, (as the strong hint for V404 Cygni at HE may exemplify) and might be detectable for especially bright outbursts, or with future gamma-ray telescopes like the Cherenkov Telescope Array \citep{paredes13,cta19}.

It should be noted that observations in the $100 - 1000$~GHz band during the state transitions would be very useful to constrain the non-thermal emitter properties by means of  establishing the transition frequency between the optically thin and optically thick synchrotron regimes, as well as determining whether non-thermal particles are accelerated up to at least a few hundred MeV or not. Upcoming MeV missions, like the All-sky Medium Energy Gamma-ray Observatory \citep[AMEGO;][]{mcenery19} or e-ASTROGRAM \citep{deangelis17}, will also be useful to bridge the $1-100$~MeV gap in observations of outbursts in X-ray binaries.


\section*{Author contribution statement}

All authors have read and approved the manuscript and contributed in one or several of the following ways: 
design, construction, maintenance and operation of the instruments used to acquire the data; preparation and/or evaluation of the observation proposals; data acquisition, processing, calibration and/or reduction; 
production of analysis tools and/or related Monte Carlo simulations. 
Notable individual contributions in alphabetical order from MAGIC collaboration members include:
J. Hoang - MAGIC analysis cross-check;
E. Molina - project leadership, coordination of the MAGIC observations, MAGIC data analysis, theoretical interpretation, paper drafting and editing;
M. Rib\'o - triggering and initial coordination of the MAGIC observations, establishment of the agreement to work with the other collaborations and contact with the multiwavelength community.
Contributions from authors outside of the MAGIC, H.E.S.S. and VERITAS collaborations include:
V. Bosch-Ramon - theoretical interpretation;
C. Celma - TJO data analysis, \textit{Swift}/UVOT data analysis;
M. Linares - TJO data analysis, \textit{Swift}/UVOT data analysis;
D.M. Russell - TJO data analysis, \textit{Swift}/UVOT data analysis;
G. Sala - TJO data analysis.

Given the wide array of contributions from many individuals over decades to bring H.E.S.S. and VERITAS to fruition -- including to design the instruments, to build them, to maintain them, to operate them, and to develop analysis software -- in addition to the efforts needed to obtain, analyse, and interpret the data for a particular scientific study, it is the policy of both the H.E.S.S. and the VERITAS Collaborations not to attempt to summarise the contributions of individual members. 


\section*{Acknowledgements}

MAGIC acknowledgements: MAGIC would like to thank the Instituto de Astrof\'{\i}sica de Canarias for the excellent working conditions at the Observatorio del Roque de los Muchachos in La Palma. The financial support of the German BMBF, MPG and HGF; the Italian INFN and INAF; the Swiss National Fund SNF; the grants PID2019-104114RB-C31, PID2019-104114RB-C32, PID2019-104114RB-C33, PID2019-105510GB-C31, PID2019-107847RB-C41, PID2019-107847RB-C42, PID2019-107847RB-C44, PID2019-107988GB-C22 funded by MCIN/AEI/ 10.13039/501100011033; the Indian Department of Atomic Energy; the Japanese ICRR, the University of Tokyo, JSPS, and MEXT; the Bulgarian Ministry of Education and Science, National RI Roadmap Project DO1-400/18.12.2020 and the Academy of Finland grant nr. 320045 is gratefully acknowledged. This work was also been supported by Centros de Excelencia ``Severo Ochoa'' y Unidades ``Mar\'{\i}a de Maeztu'' program of the MCIN/AEI/ 10.13039/501100011033 (SEV-2016-0588, SEV-2017-0709, CEX2019-000920-S, CEX2019-000918-M, MDM-2015-0509-18-2) and by the CERCA institution of the Generalitat de Catalunya; by the Croatian Science Foundation (HrZZ) Project IP-2016-06-9782 and the University of Rijeka Project uniri-prirod-18-48; by the DFG Collaborative Research Centers SFB1491 and SFB876/C3; the Polish Ministry Of Education and Science grant No. 2021/WK/08; and by the Brazilian MCTIC, CNPq and FAPERJ.

\vskip 3mm

HESS acknowledgements: The support of the Namibian authorities and of the University of Namibia in facilitating the construction and operation of H.E.S.S. is gratefully acknowledged, as is the support by the German Ministry for Education and Research (BMBF), the Max Planck Society, the German Research Foundation (DFG), the Helmholtz Association, the Alexander von Humboldt Foundation, the French Ministry of Higher Education, Research and Innovation, the Centre National de la Recherche Scientifique (CNRS/IN2P3 and CNRS/INSU), the Commissariat \`a l'\'energie atomique et aux \'energies alternatives (CEA), the U.K. Science and Technology Facilities Council (STFC), the Irish Research Council (IRC) and the Science Foundation Ireland (SFI), the Knut and Alice Wallenberg Foundation, the Polish Ministry of Education and Science, agreement no. 2021/WK/06, the South African Department of Science and Technology and National Research Foundation, the University of Namibia, the National Commission on Research, Science \& Technology of Namibia (NCRST), the Austrian Federal Ministry of Education, Science and Research and the Austrian Science Fund (FWF), the Australian Research Council (ARC), the Japan Society for the Promotion of Science, the University of Amsterdam and the Science Committee of Armenia grant 21AG-1C085. We appreciate the excellent work of the technical support staff in Berlin, Zeuthen, Heidelberg, Palaiseau, Paris, Saclay, T\"ubingen and in Namibia in the construction and operation of the equipment. This work benefited from services provided by the H.E.S.S. Virtual Organisation, supported by the national resource providers of the EGI Federation.

VERITAS acknowledgements: VERITAS is supported by grants from the U.S. Department of Energy Office of Science, the U.S. National Science Foundation and the Smithsonian Institution, by NSERC in Canada, and by the Helmholtz Association in Germany. This research used resources provided by the Open Science Grid, which is supported by the National Science Foundation and the U.S. Department of Energy's Office of Science, and resources of the National Energy Research Scientific Computing Center (NERSC), a U.S. Department of Energy Office of Science User Facility operated under Contract No. DE-AC02-05CH11231. We acknowledge the excellent work of the technical support staff at the Fred Lawrence Whipple Observatory and at the collaborating institutions in the construction and operation of the instrument.

We would also like to thank the referee for his/her constructive and useful comments, which were helpful to improve the manuscript. E. Molina acknowledges support from MCIN through grant BES-2016-076342. V. Bosch-Ramon acknowledges financial support from the State Agency for Research of the Spanish Ministry of Science and Innovation under grant PID2019-105510GB-C31 and through the “Unit of Excellence María de Maeztu 2020-2023” award to the Institute of Cosmos Sciences (CEX2019-000918-M). V. Bosch-Ramon is Correspondent Researcher of CONICET, Argentina, at the IAR. M. Linares has received funding from the European Research Council (ERC) under the European Union’s Horizon 2020 research and innovation programme (grant agreement No. 101002352). G. Sala acknowledges support from the Spanish MINECO grant PID2020-117252GB-I00.

We acknowledge the {\it Fermi}-LAT collaboration for making available the data and the analysis tools used in this work. This research has made use of MAXI data provided by RIKEN, JAXA and the MAXI team. This work also made use of data supplied by the UK Swift Science Data Centre at the University of Leicester. We also used data from the The Joan Oró Telescope (TJO) of the Montsec Astronomical Observatory (OAdM), which is owned by the Catalan Government and operated by the Institute for Space Studies of Catalonia (IEEC). We would also like to thank Sera Markoff, Phil Uttley, and our colleagues from the X-ray, optical/infrared, and radio communities, for the very fruitful exchanges we had on MAXI~J1820+070 during the observational campaign.


\section*{Data availability}

Part of the data underlying this article are available in the article and in its online supplementary material. In particular, MAXI/GSC data were obtained from \url{http://maxi.riken.jp/star_data/J1820+071/J1820+071.html}; \textit{Swift}/BAT data were downloaded from \url{https://swift.gsfc.nasa.gov/results/transients/weak/MAXIJ1820p070}; NICER data were retrieved from \url{https://heasarc.gsfc.nasa.gov/docs/archive.html}, and the corresponding calibration data were downloaded from \url{https://heasarc.gsfc.nasa.gov/docs/heasarc/caldb/data/nicer/xti/index/cif_nicer_xti_20200722.html}; and \textit{Fermi}-LAT data were downloaded from \url{https://fermi.gsfc.nasa.gov/cgi-bin/ssc/LAT/LATDataQuery.cgi}. The rest of the data will be shared on reasonable request to the corresponding authors.


\FloatBarrier
\bibliographystyle{mnras}
\bibliography{references}


\appendix

\section{Observation dates}\label{sec:tables}

\begin{table}
    \begin{center}
    \caption{Dates when H.E.S.S., MAGIC and/or VERITAS observations of MAXI~J1820+070 were performed. Only dates with surviving data after quality cuts are shown. These are also depicted in Fig.~\ref{fig:HR}.}
    \begin{tabular}{c c c c c}
    \hline \hline
    Date    & Date          & H.E.S.S.      & MAGIC         & VERITAS       \\
    {[MJD]} & [Gregorian]   &               &               &               \\
    \hline
    58197   & 20 Mar. 2018  &               &               & $\checkmark$  \\
    58199   & 22 Mar. 2018  & $\checkmark$  &               &               \\
    58200   & 23 Mar. 2018  & $\checkmark$  &               &               \\	 
    58201   & 24 Mar. 2018  & $\checkmark$  & $\checkmark$  &               \\
    58202   & 25 Mar. 2018  & $\checkmark$  &               &               \\
    58204   & 27 Mar. 2018  & $\checkmark$  & $\checkmark$  &               \\
    58220   & 12 Apr. 2018  &               &               & $\checkmark$  \\
    58221   & 13 Apr. 2018  &               &               & $\checkmark$  \\
    58222   & 14 Apr. 2018  &               &               & $\checkmark$  \\
    58223   & 15 Apr. 2018  &               &               & $\checkmark$  \\
    58224   & 16 Apr. 2018  &               &               & $\checkmark$  \\
    58227   & 19 Apr. 2018  & $\checkmark$  &               &               \\
    58229   & 21 Apr. 2018  & $\checkmark$  &               & $\checkmark$  \\
    58230   & 22 Apr. 2018  & $\checkmark$  &               &               \\
    58231   & 23 Apr. 2018  & $\checkmark$  &               &               \\
    58232   & 24 Apr. 2018  & $\checkmark$  &               &               \\
    58233   & 25 Apr. 2018  & $\checkmark$  &               &               \\
    58234   & 26 Apr. 2018  & $\checkmark$  &               &               \\
    58235   & 27 Apr. 2018  & $\checkmark$  &               &               \\
    58276   & 7 Jun. 2018   & $\checkmark$  &               & $\checkmark$  \\
    58277   & 8 Jun. 2018   & $\checkmark$  &               & $\checkmark$  \\
    58278   & 9 Jun. 2018   & $\checkmark$  &               & $\checkmark$  \\
    58279   & 10 Jun. 2018  &               & $\checkmark$  & $\checkmark$  \\
    58280   & 11 Jun. 2018  & $\checkmark$  & $\checkmark$  & $\checkmark$  \\
    58281   & 12 Jun. 2018  & $\checkmark$  & $\checkmark$  & $\checkmark$  \\
    58282   & 13 Jun. 2018  &               & $\checkmark$  &               \\
    58283   & 14 Jun. 2018  &               & $\checkmark$  &               \\
    58284   & 15 Jun. 2018  & $\checkmark$  & $\checkmark$  &               \\
    58287   & 18 Jun. 2018  &               & $\checkmark$  &               \\
    58288   & 19 Jun. 2018  &               & $\checkmark$  &               \\
    58291   & 22 Jun. 2018  &               & $\checkmark$  &               \\
    58306   & 7 Jul. 2018   &               & $\checkmark$  &               \\
    58307   & 8 Jul. 2018   & $\checkmark$  & $\checkmark$  &               \\
    58309   & 10 Jul. 2018  & $\checkmark$  &               &               \\
    58313   & 14 Jul. 2018  & $\checkmark$  &               &               \\
    58314   & 15 Jul. 2018  & $\checkmark$  &               &               \\
    58317   & 18 Jul. 2018  & $\checkmark$  &               &               \\
    58389   & 28 Sep. 2018  & $\checkmark$  &               &               \\
    58390   & 29 Sep. 2018  & $\checkmark$  & $\checkmark$  &               \\
    58391   & 30 Sep. 2018  & $\checkmark$  & $\checkmark$  &               \\
    58392   & 1 Oct. 2018   & $\checkmark$  & $\checkmark$  &               \\
    \hline
    \end{tabular}
    \label{tab:dates}
    \end{center}
\end{table}


\section{VHE gamma-ray flux computation}\label{sec:fluxComputation}

For each experiment and energy bin, the low-level data analysis yields: the number of gamma-ray events recorded in the direction of the source (ON region) and in control regions with only background events (OFF regions), $N_{\rm on}$ and $N_{\rm off}$, respectively; the exposure ratio of the OFF to ON regions, $\tau$; the effective observation time of the source after data quality cuts, $t_{\rm eff}$; and the effective collection area averaged over the considered energy interval, $<\!A_{\rm eff}\!>$. We assume a power-law distribution for the gamma rays coming from the source, i.e. d$N/$d$\varepsilon = K \varepsilon^{-\alpha}$, where $N$ is the number of gamma-ray photons, $K$ is the flux normalisation constant, $\varepsilon$ is the gamma-ray energy, and $\alpha$ is the spectral index. With this, the expected number of gamma rays coming from the source in the energy interval $[\varepsilon_{\rm min}$ , $\varepsilon_{\rm max}]$ can be expressed as
    \begin{linenomath}\begin{equation}\label{expectedSignal}
    \mu = t_{\rm eff} \int_{\varepsilon_{\rm min}}^{\varepsilon_{\rm max}} A_{\rm eff}(\varepsilon) \frac{{\rm d}N}{{\rm d}\varepsilon} {\rm d}\varepsilon
    = K <\!A_{\rm eff}\!> t_{\rm eff} \frac{\varepsilon_{\rm min}^{1-\alpha} - \varepsilon_{\rm max}^{1-\alpha}}{\alpha - 1} \ .
    \end{equation}\end{linenomath}
In order to obtain the range of values of $K$ compatible with the observed quantities, the value of $\alpha$ is fixed, and a maximum likelihood method is performed as described by \cite{rolke05}. We define a Poissonian likelihood function as
    \begin{linenomath}\begin{equation}\label{likeFunction}
    L = \frac{(\epsilon \mu + b)^{N_{\rm on}}}{N_{\rm on}!} \times
        \frac{(\tau b)^{N_{\rm off}}}{N_{\rm off}!} \times
        \frac{1}{\sigma_{\epsilon} \sqrt{2\pi}} \exp\left[
        -\frac{1}{2} \left( \frac{\epsilon - \epsilon_0}{\sigma_{\epsilon}} \right)^2 \right] \ ,
    \end{equation}\end{linenomath}
where the terms correspond, from left to right, to the statistical distributions of a Poissonian signal, a Poissonian background, and a detection efficiency with a Gaussian uncertainty. The factors $b$ and $\epsilon$, which are treated as nuisance parameters, are the expected number of background events in the signal region, and the expected detector efficiency, respectively. The parameters $\epsilon_0$ and $\sigma_\epsilon$ are the estimates for the efficiency and its standard deviation, respectively. Fixing $\epsilon_0 = 1$ allows us to account for the relative systematic uncertainty of the instrument by equating it to the value of $\sigma_\epsilon$.

With the likelihood function defined, we find the values $\hat{K}$, $\hat{b}$ and $\hat{\epsilon}$ that maximise $L$, which can be obtained analytically for the likelihood function expressed in Eq.~\eqref{likeFunction}. The null hypothesis $K = K_0$ is then tested versus the alternative hypothesis $K \ne K_0$ through a likelihood ratio test statistic:
    \begin{linenomath}\begin{equation}\label{likeRatio}
    \lambda = \frac{L(K_0, \hat{b}(K_0), \hat{\epsilon}(K_0))}{L(\hat{K}, \hat{b}, \hat{\epsilon})} \ ,
    \end{equation}\end{linenomath}
where $\hat{b}(K_0)$ and $\hat{\epsilon}(K_0)$ are the values that maximise $L$ for a given $K_0$. According to the Wilks theorem \citep{wilks38}, under the null hypothesis the distribution of the quantity $-2 \ln \lambda$ converges to a $\chi^2$ distribution with 1 degree of freedom for large enough statistics. This allows us to find the range of $K_0$ compatible with the observations in the energy bin $[\varepsilon_{\rm min}$ , $\varepsilon_{\rm max}]$, i.e., being $n = \sqrt{-2 \ln \lambda}$, the null hypothesis is excluded at a $n \sigma$ level. Finally, the upper end of this range of $K_0$ is translated to an upper limit in flux (with a $n \sigma$ confidence level) through the assumed spectral shape.

The method explained above works for the flux computation of each individual experiment. In order to merge H.E.S.S., MAGIC and VERITAS data into a single flux measurement, a joint likelihood function is defined as the product of the individual likelihoods defined in Eq.~\eqref{likeFunction}:
    \begin{linenomath}\begin{equation}\label{likeJoint}
    L_{\rm tot} = \prod_{i = 1}^{3} L_i(N_{{\rm on,}i}, N_{{\rm off,}i}, \tau_i, t_{{\rm eff,}i}, A_{{\rm eff,}i}) \ .
    \end{equation}\end{linenomath}
Given that the data of each experiment are independent from each other, the maximisation procedure can be done individually for each instrument. Therefore, with the joint likelihood defined in Eq.~\eqref{likeJoint}, we have
    \begin{linenomath}\begin{equation}\label{profJoint}
    -2 \ln \lambda_{\rm tot} = \sum_{i = 1}^{3} -2 \ln \lambda_{\rm i}
    \end{equation}\end{linenomath}
for each value of $K_0$, and the null hypothesis is rejected or not according to the same criteria as in the individual case.

\bsp
\vskip 5mm
\input{affiliations}

\label{lastpage}

\end{document}

%% file: author_list.tex
\author[H.~Abe~et~al.]{\parbox{\textwidth}{\centering\scriptsize{
H.~Abe$^{1}$,
S.~Abe$^{1}$,
V.~A.~Acciari$^{2}$,
T.~Aniello$^{3}$,
S.~Ansoldi$^{4,45}$,
L.~A.~Antonelli$^{3}$,
A.~Arbet Engels$^{5}$,
C.~Arcaro$^{6}$,
M.~Artero$^{7}$,
K.~Asano$^{1}$
% ...
\\
(The MAGIC Collaboration)
\\
\vspace{1.5mm}
H.~Abdalla$^{53}$,
F.~Aharonian$^{54,55}$,
F.~Ait~Benkhali$^{56}$,
E.~O.~Ang\"uner$^{57}$,
H.~Ashkar$^{58}$,
M.~Backes$^{53,59}$,
V.~Baghmanyan$^{60}$,
V.~Barbosa~Martins$^{61}$,
R.~Batzofin$^{62}$,
Y.~Becherini$^{63,64}$
% ...
\\
(The VERITAS Collaboration)
\\
\vspace{1.5mm}
V.~Bosch-Ramon$^{17}$,
C.~Celma$^{120}$,
M.~Linares$^{121,120}$,
D.~M.~Russell$^{122}$,
G.~Sala$^{120,123}$
\\
\vspace{2mm}
\textit{Affiliations are provided at the end of the paper}
\vspace{-2mm}
}}}

%% file: affiliations.tex
MAGIC Collaboration:\\
$^{1}$ {Japanese MAGIC Group: Institute for Cosmic Ray Research (ICRR), The University of Tokyo, Kashiwa, 277-8582 Chiba, Japan} \\
$^{2}$ {Instituto de Astrof\'isica de Canarias and Dpto. de  Astrof\'isica, Universidad de La Laguna, E-38200, La Laguna, Tenerife, Spain} \\
$^{3}$ {National Institute for Astrophysics (INAF), I-00136 Rome, Italy} \\
$^{4}$ {Universit\`a di Udine and INFN Trieste, I-33100 Udine, Italy} \\
$^{5}$ {Max-Planck-Institut f\"ur Physik, D-80805 M\"unchen, Germany} \\
$^{6}$ {Universit\`a di Padova and INFN, I-35131 Padova, Italy} \\
$^{7}$ {Institut de F\'isica d'Altes Energies (IFAE), The Barcelona Institute of Science and Technology (BIST), E-08193 Bellaterra (Barcelona), Spain} \\
$^{8}$ {Technische Universit\"at Dortmund, D-44221 Dortmund, Germany} \\
$^{9}$ {Croatian MAGIC Group: University of Zagreb, Faculty of Electrical Engineering and Computing (FER), 10000 Zagreb, Croatia} \\
$^{10}$ {IPARCOS Institute and EMFTEL Department, Universidad Complutense de Madrid, E-28040 Madrid, Spain} \\
$^{11}$ {Centro Brasileiro de Pesquisas F\'isicas (CBPF), 22290-180 URCA, Rio de Janeiro (RJ), Brazil} \\
$^{12}$ {University of Lodz, Faculty of Physics and Applied Informatics, Department of Astrophysics, 90-236 Lodz, Poland} \\
$^{13}$ {Instituto de Astrof\'isica de Andaluc\'ia-CSIC, Glorieta de la Astronom\'ia s/n, 18008, Granada, Spain} \\
$^{14}$ {Deutsches Elektronen-Synchrotron (DESY), D-15738 Zeuthen, Germany} \\
$^{15}$ {ETH Z\"urich, CH-8093 Z\"urich, Switzerland} \\
$^{16}$ {Universit\`a di Pisa and INFN Pisa, I-56126 Pisa, Italy} \\
$^{17}$ {Universitat de Barcelona, ICCUB, IEEC-UB, E-08028 Barcelona, Spain} \\
$^{18}$ {Armenian MAGIC Group: A. Alikhanyan National Science Laboratory, 0036 Yerevan, Armenia} \\
$^{19}$ {Centro de Investigaciones Energ\'eticas, Medioambientales y Tecnol\'ogicas, E-28040 Madrid, Spain} \\
$^{20}$ {Department for Physics and Technology, University of Bergen, Norway} \\
$^{21}$ {INFN MAGIC Group: INFN Sezione di Catania and Dipartimento di Fisica e Astronomia, University of Catania, I-95123 Catania, Italy} \\
$^{22}$ {INFN MAGIC Group: INFN Sezione di Torino and Universit\`a degli Studi di Torino, I-10125 Torino, Italy} \\
$^{23}$ {INFN MAGIC Group: INFN Sezione di Bari and Dipartimento Interateneo di Fisica dell'Universit\`a e del Politecnico di Bari, I-70125 Bari, Italy} \\
$^{24}$ {Croatian MAGIC Group: University of Rijeka, Faculty of Physics, 51000 Rijeka, Croatia} \\
$^{25}$ {Universit\"at W\"urzburg, D-97074 W\"urzburg, Germany} \\
$^{26}$ {University of Geneva, Chemin d'Ecogia 16, CH-1290 Versoix, Switzerland} \\
$^{27}$ {Finnish MAGIC Group: Finnish Centre for Astronomy with ESO, University of Turku, FI-20014 Turku, Finland} \\
$^{28}$ {Departament de F\'isica, and CERES-IEEC, Universitat Aut\`onoma de Barcelona, E-08193 Bellaterra, Spain} \\
$^{29}$ {Japanese MAGIC Group: Physics Program, Graduate School of Advanced Science and Engineering, Hiroshima University, 739-8526 Hiroshima, Japan} \\
$^{30}$ {Armenian MAGIC Group: ICRANet-Armenia at NAS RA, 0019 Yerevan, Armenia} \\
$^{31}$ {Croatian MAGIC Group: University of Split, Faculty of Electrical Engineering, Mechanical Engineering and Naval Architecture (FESB), 21000 Split, Croatia} \\
$^{32}$ {Croatian MAGIC Group: Josip Juraj Strossmayer University of Osijek, Department of Physics, 31000 Osijek, Croatia} \\
$^{33}$ {Japanese MAGIC Group: Department of Physics, Tokai University, Hiratsuka, 259-1292 Kanagawa, Japan} \\
$^{34}$ {Universit\`a di Siena and INFN Pisa, I-53100 Siena, Italy} \\
$^{35}$ {Saha Institute of Nuclear Physics, A CI of Homi Bhabha National Institute, Kolkata 700064, West Bengal, India} \\
$^{36}$ {Inst. for Nucl. Research and Nucl. Energy, Bulgarian Academy of Sciences, BG-1784 Sofia, Bulgaria} \\
$^{37}$ {Japanese MAGIC Group: Department of Physics, Yamagata University, Yamagata 990-8560, Japan} \\
$^{38}$ {Finnish MAGIC Group: Space Physics and Astronomy Research Unit, University of Oulu, FI-90014 Oulu, Finland} \\
$^{39}$ {Japanese MAGIC Group: Department of Physics, Kyoto University, 606-8502 Kyoto, Japan} \\
$^{40}$ {Japanese MAGIC Group: Institute for Space-Earth Environmental Research and Kobayashi-Maskawa Institute for the Origin of Particles and the Universe, Nagoya University, 464-6801 Nagoya, Japan} \\
$^{41}$ {Croatian MAGIC Group: Ru\dj{}er Bo\v{s}kovi\'c Institute, 10000 Zagreb, Croatia} \\
$^{42}$ {INFN MAGIC Group: INFN Sezione di Perugia, I-06123 Perugia, Italy} \\
$^{43}$ {INFN MAGIC Group: INFN Roma Tor Vergata, I-00133 Roma, Italy} \\
$^{44}$ {Japanese MAGIC Group: Department of Physics, Konan University, Kobe, Hyogo 658-8501, Japan} \\
$^{45}$ {also at International Center for Relativistic Astrophysics (ICRA), Rome, Italy} \\
$^{46}$ {now at University of Innsbruck, Institute for Astro and Particle Physics} \\
$^{47}$ {also at Port d'Informaci\'o Cient\'ifica (PIC), E-08193 Bellaterra (Barcelona), Spain} \\
$^{48}$ {now at Ruhr-Universit\"at Bochum, Fakult\"at f\"ur Physik und Astronomie, Astronomisches Institut (AIRUB), 44801 Bochum, Germany} \\
$^{49}$ {also at University of Innsbruck, Institute for Astro- and Particle Physics} \\
$^{50}$ {also at Dipartimento di Fisica, Universit\`a di Trieste, I-34127 Trieste, Italy} \\
$^{51}$ {also at University of Lodz, Faculty of Physics and Applied Informatics, Department of Astrophysics, 90-236 Lodz, Poland} \\
$^{52}$ {also at INAF Trieste and Dept. of Physics and Astronomy, University of Bologna, Bologna, Italy} \\

H.E.S.S. Collaboration:\\
$^{53}$ University of Namibia, Department of Physics, Private Bag 13301, Windhoek 10005, Namibia \\
$^{54}$ Dublin Institute for Advanced Studies, 31 Fitzwilliam Place, Dublin 2, Ireland \\
$^{55}$ Max-Planck-Institut f\"ur Kernphysik, P.O. Box 103980, D 69029 Heidelberg, Germany \\
$^{56}$ Landessternwarte, Universit\"at Heidelberg, K\"onigstuhl, D 69117 Heidelberg, Germany \\
$^{57}$ Aix Marseille Universit\'e, CNRS/IN2P3, CPPM, Marseille, France \\
$^{58}$ Laboratoire Leprince-Ringuet, \'Ecole Polytechnique, CNRS, Institut Polytechnique de Paris, F-91128 Palaiseau, France \\
$^{59}$ Centre for Space Research, North-West University, Potchefstroom 2520, South Africa \\
$^{60}$ Instytut Fizyki J\c{a}drowej PAN, ul. Radzikowskiego 152, 31-342 Krak{\'o}w, Poland \\
$^{61}$ DESY, D-15738 Zeuthen, Germany \\
$^{62}$ School of Physics, University of the Witwatersrand, 1 Jan Smuts Avenue, Braamfontein, Johannesburg, 2050 South Africa \\
$^{63}$ Universit\'e Paris Cit\'e, CNRS, Astroparticule et Cosmologie, F-75013 Paris, France \\
$^{64}$ Department of Physics and Electrical Engineering, Linnaeus University,  351 95 V\"axj\"o, Sweden \\
$^{65}$ Institut f\"ur Physik, Humboldt-Universit\"at zu Berlin, Newtonstr. 15, D 12489 Berlin, Germany \\
$^{66}$ Laboratoire Univers et Th\'eories, Observatoire de Paris, Universit\'e PSL, CNRS, Universit\'e Paris Cit\'e, 92190 Meudon, France \\
$^{67}$ Sorbonne Universit\'e, Universit\'e Paris Diderot, Sorbonne Paris Cit\'e, CNRS/IN2P3, Laboratoire de Physique Nucl\'eaire et de Hautes Energies, LPNHE, 4 Place Jussieu, F-75252 Paris, France \\
$^{68}$ Universit\'e Savoie Mont Blanc, CNRS, Laboratoire d'Annecy de Physique des Particules - IN2P3, 74000 Annecy, France \\
$^{69}$ IRFU, CEA, Universit\'e Paris-Saclay, F-91191 Gif-sur-Yvette, France \\
$^{70}$ Astronomical Observatory, The University of Warsaw, Al. Ujazdowskie 4, 00-478 Warsaw, Poland \\
$^{71}$ University of Oxford, Department of Physics, Denys Wilkinson Building, Keble Road, Oxford OX1 3RH, UK \\
$^{72}$ Universit\'e Bordeaux, CNRS, LP2I Bordeaux, UMR 5797, F-33170 Gradignan, France \\
$^{73}$ Institut f\"ur Physik und Astronomie, Universit\"at Potsdam,  Karl-Liebknecht-Strasse 24/25, D 14476 Potsdam, Germany \\
$^{74}$ Friedrich-Alexander-Universit\"at Erlangen-N\"urnberg, Erlangen Centre for Astroparticle Physics, Erwin-Rommel-Str. 1, D 91058 Erlangen, Germany \\
$^{75}$ Institut f\"ur Astro- und Teilchenphysik, Leopold-Franzens-Universit\"at Innsbruck, A-6020 Innsbruck, Austria \\
$^{76}$ Universit\"at Hamburg, Institut f\"ur Experimentalphysik, Luruper Chaussee 149, D 22761 Hamburg, Germany \\
$^{77}$ Obserwatorium Astronomiczne, Uniwersytet Jagiello{\'n}ski, ul. Orla 171, 30-244 Krak{\'o}w, Poland \\
$^{78}$ Institute of Astronomy, Faculty of Physics, Astronomy and Informatics, Nicolaus Copernicus University,  Grudziadzka 5, 87-100 Torun, Poland \\
$^{79}$ Nicolaus Copernicus Astronomical Center, Polish Academy of Sciences, ul. Bartycka 18, 00-716 Warsaw, Poland \\
$^{80}$ Institut f\"ur Astronomie und Astrophysik, Universit\"at T\"ubingen, Sand 1, D 72076 T\"ubingen, Germany \\
$^{81}$ Department of Physics and Astronomy, The University of Leicester, University Road, Leicester, LE1 7RH, United Kingdom \\
$^{82}$ GRAPPA, Anton Pannekoek Institute for Astronomy, University of Amsterdam,  Science Park 904, 1098 XH Amsterdam, The Netherlands \\
$^{83}$ Laboratoire Univers et Particules de Montpellier, Universit\'e Montpellier, CNRS/IN2P3,  CC 72, Place Eug\`ene Bataillon, F-34095 Montpellier Cedex 5, France \\
$^{84}$ School of Physical Sciences, University of Adelaide, Adelaide 5005, Australia \\
$^{85}$ Yerevan Physics Institute, 2 Alikhanian Brothers St., 375036 Yerevan, Armenia \\
$^{86}$ Department of Physics, Konan University, 8-9-1 Okamoto, Higashinada, Kobe, Hyogo 658-8501, Japan \\
$^{87}$ Kavli Institute for the Physics and Mathematics of the Universe (WPI), The University of Tokyo Institutes for Advanced Study (UTIAS), The University of Tokyo, 5-1-5 Kashiwa-no-Ha, Kashiwa, Chiba, 277-8583, Japan \\
$^{88}$ RIKEN, 2-1 Hirosawa, Wako, Saitama 351-0198, Japan \\
$^{89}$ Department of Physics, Rikkyo University, 3-34-1 Nishi-Ikebukuro, Toshima-ku, Tokyo 171-8501, Japan \\
$^{90}$ Department of Physics, University of the Free State,  PO Box 339, Bloemfontein 9300, South Africa \\
$^{91}$ Kapteyn Astronomical Institute, University of Groningen, Landleven 12, 9747 AD Groningen, The Netherlands \\

VERITAS Collaboration:\\
$^{92}${Department of Physics and Astronomy, University of Alabama, Tuscaloosa, AL 35487, USA}\\
$^{93}${Physics Department, Columbia University, New York, NY 10027, USA}\\
$^{94}${DESY, Platanenallee 6, 15738 Zeuthen, Germany}\\
$^{95}${Center for Astrophysics $|$ Harvard \& Smithsonian, Cambridge, MA 02138, USA}\\
$^{96}${Department of Physics and Astronomy, Barnard College, Columbia University, NY 10027, USA}\\
$^{97}${Physics Department, California Polytechnic State University, San Luis Obispo, CA 94307, USA}\\
$^{98}${Department of Physics and Astronomy, Iowa State University, Ames, IA 50011, USA}\\
$^{99}${Department of Physics, Washington University, St. Louis, MO 63130, USA}\\
$^{100}${Department of Astronomy and Astrophysics, 525 Davey Lab, Pennsylvania State University, University Park, PA 16802, USA}\\
$^{101}${Department of Physics and Astronomy, Purdue University, West Lafayette, IN 47907, USA}\\
$^{102}${Department of Physics and Astronomy and the Bartol Research Institute, University of Delaware, Newark, DE 19716, USA}\\
$^{103}${School of Physics and Astronomy, University of Minnesota, Minneapolis, MN 55455, USA}\\
$^{104}${Department of Physics, California State University - East Bay, Hayward, CA 94542, USA}\\
$^{105}${School of Physics and Center for Relativistic Astrophysics, Georgia Institute of Technology, 837 State Street NW, Atlanta, GA 30332-0430}\\
$^{106}${Santa Cruz Institute for Particle Physics and Department of Physics, University of California, Santa Cruz, CA 95064, USA}\\
$^{107}${Department of Physics and Astronomy, University of Utah, Salt Lake City, UT 84112, USA}\\
$^{108}${Department of Physics and Astronomy, University of Iowa, Van Allen Hall, Iowa City, IA 52242, USA}\\
$^{109}${Department of Physics and Astronomy, DePauw University, Greencastle, IN 46135-0037, USA}\\
$^{110}${Physics Department, McGill University, Montreal, QC H3A 2T8, Canada}\\
$^{111}${School of Physics, National University of Ireland Galway, University Road, Galway, Ireland}\\
$^{112}${School of Physics, University College Dublin, Belfield, Dublin 4, Ireland}\\
$^{113}${Department of Physics and Astronomy, Ball State University, Muncie, IN 47306, USA}\\
$^{114}${Department of Physics, Anderson University, 1100 East 5th Street, Anderson, IN 46012}\\
$^{115}${Department of Physics and Astronomy, University of California, Los Angeles, CA 90095, USA}\\
$^{116}${Department of Physics, Queen's University, Kingston, ON K7L 3N6, Canada}\\
$^{117}${Institute of Physics and Astronomy, University of Potsdam, 14476 Potsdam-Golm, Germany}\\
$^{118}${Department of Physical Sciences, Munster Technological University, Bishopstown, Cork, T12 P928, Ireland}\\
$^{119}${Department of Physics, Indiana University-Purdue University Indianapolis, Indianapolis, IN 46202, USA}\\

$^{120}$ Departament de F{\'i}sica, Universitat Polit{\`e}cnica de Catalunya (UPC), E-08019 Barcelona, Spain\\
$^{121}$ Institutt for Fysikk, Norwegian University of Science and Technology, Trondheim, Norway\\
$^{122}$ Center for Astro, Particle and Planetary Physics, New York University Abu Dhabi, PO Box 129188, Abu Dhabi, UAE\\
$^{123}$ Institut d'Estudis Espacials de Catalunya (IEEC), E-08034 Barcelona, Spain